\newif\ifarxiv\arxivtrue
\arxivfalse    
\ifarxiv\pdfoutput=1\fi
{\def\usepackage{ws-procs9x6}}
\documentclass[12pt,a4paper]{article}

\usepackage{amsmath,amssymb}
\usepackage[bookmarks=true,hyperfigures=true]{hyperref}
\usepackage{graphicx}
\usepackage[nosort]{cite}
\usepackage[bulletsep]{collref}
\usepackage{fixmath}
\usepackage{verbatim}

\usepackage[a4paper,text={160mm,247mm},centering]{geometry}

\def\showkeysrefformat#1{{\normalfont\tiny\ttfamily#1}}
\makeatletter
\def\SK@@ref#1>#2\SK@{%
 {\@inlabelfalse\leavevmode\vbox to\z@{%
 \vss\SK@refcolor\rlap{\vrule\raise .75em%
  \hbox{\showkeysrefformat{#2}}}}}}
\makeatother

\allowdisplaybreaks[3]

\numberwithin{equation}{section}

\usepackage[font=small,labelfont=bf,width=0.85\textwidth]{caption}

\expandafter\def\expandafter\bfseries\expandafter{\bfseries\ifmmode\else\boldmath\fi}
\expandafter\def\expandafter\mdseries\expandafter{\mdseries\ifmmode\else\unboldmath\fi}
\expandafter\def\expandafter\normalfont\expandafter{\normalfont\ifmmode\else\unboldmath\fi}

\RequirePackage{verbatim}

\makeatletter
\newwrite\bibinl@out
\newenvironment{bibtex}[1]{%
  \immediate\openout\bibinl@out #1.bib
  \immediate\write\bibinl@out{\@percentchar generated from `\jobname' starting line \the\inputlineno^^J}%
  \def\verbatim@processline{\immediate\write\bibinl@out{\the\verbatim@line}}%
  \@bsphack\let\do\@makeother\dospecials\catcode`\^^M\active\verbatim@start
}%
{\immediate\closeout\bibinl@out\@esphack}
\makeatother

\def\[{\begin{equation}}
\def\]{\end{equation}}

\providecommand{\href}[2]{#2}

\makeatletter
\def\mr@ignsp#1 {\ifx\:#1\@empty\else #1\expandafter\mr@ignsp\fi}%
\newcommand{\multiref}[1]{\begingroup
\xdef\mr@no@sparg{\expandafter\mr@ignsp#1 \: }%
\def\mr@comma{}%
\@for\mr@refs:=\mr@no@sparg\do{\mr@comma\def\mr@comma{,}\ref{\mr@refs}}%
\endgroup}
\renewcommand{\eqref}[1]{(\multiref{#1})}
\makeatother

\makeatletter
\newcommand{\namedref}[2]{\hyperref[#2]{#1~\ref*{#2}}}
\newcommand{\secref}{\@ifstar{\namedref{Section}}{\namedref{Sec.}}}
\newcommand{\appref}{\@ifstar{\namedref{Appendix}}{\namedref{App.}}}
\newcommand{\tabref}{\@ifstar{\namedref{Table}}{\namedref{Tab.}}}
\newcommand{\figref}{\@ifstar{\namedref{Figure}}{\namedref{Fig.}}}
\makeatother

\providecommand{\hypersetup}[1]{}
\providecommand{\texorpdfstring}[2]{#1}

\hypersetup{plainpages=false}
\hypersetup{pdfpagemode=UseNone}
\hypersetup{bookmarksnumbered=true}
\hypersetup{pdfstartview=FitH}
\hypersetup{colorlinks=false}
\hypersetup{citebordercolor={.5 1 .5}}
\hypersetup{urlbordercolor={.5 1 1}}
\hypersetup{linkbordercolor={1 .7 .7}}

\makeatletter
\let\@keywords\@empty
\let\@subject\@empty
\providecommand{\keywords}[1]{\gdef\@keywords{#1}}
\providecommand{\subject}[1]{\gdef\@subject{#1}}
\def\thetitle{\@title}
\def\theauthor{\@author}
\def\thesubject{\@subject}
\def\thedate{\@date}
\def\thekeywords{\@keywords}
\makeatother
\AtBeginDocument{
\hypersetup{pdftitle={\thetitle}}%
\hypersetup{pdfauthor={\theauthor}}%
\hypersetup{pdfsubject={\thesubject}}%
\hypersetup{pdfkeywords={\thekeywords}}%
}

\providecommand{\genfrac}[6]{\left#1{{#5}\above#3{#6}}\right#2}
\newcommand{\sfrac}[2]{{\textstyle\frac{#1}{#2}}}
\newcommand{\half}{\sfrac{1}{2}}
\newcommand{\ihalf}{\sfrac{i}{2}}
\newcommand{\alg}[1]{\mathfrak{#1}}

\newcommand{\fracchoose}{\genfrac{(}{)}{0pt}{}}

\newcommand{\kappaR}{}

\newcommand{\oper}[1]{\mathrm{#1}}
\newcommand{\opkern}[1]{\mathcal{#1}}
\newcommand{\oplocal}[1]{[#1]}
\newcommand{\opboost}[1]{\mathcal{B}[#1]}
\newcommand{\opbilocal}[2]{[#1|#2]}

\newcommand{\Left}{-}
\newcommand{\Right}{+}
\newcommand{\LeftRight}{\mp}

\RequirePackage{amsmath}
\makeatletter
\newcommand{\brk@ord}[1]{{\def\brk@temp{#1}\def\brk@dot{.}\ifx\brk@temp\brk@dot\else #1\fi}}
\newcommand{\brk@ordl}{\mathopen\brk@ord}
\newcommand{\brk@ordr}{\mathclose\brk@ord}
\newcommand{\brk@ordm}{\mathrel\brk@ord}
\newcommand{\brk@var}{\brk@ord}
\newcommand{\brk@varl}{\left}
\newcommand{\brk@varr}{\right}
\newcommand{\brk@varm}{\mathrel\brk@var}
\newcommand{\brk@altname}[3]{\expandafter\def\csname#2\expandafter\@gobble\string#1\endcsname{#1[#3]}}
\newcommand{\brk@usearg}[3]{%
  \def\brk@star{*}\def\brk@blank{}\def\brk@arg{#1}%
  \ifx\brk@arg\brk@blank\def\brk@arg{brk@ord}\fi%
  \ifx\brk@arg\brk@star\def\brk@arg{brk@var}\fi%
  \csname\brk@arg #2\endcsname#3}

\newcommand{\DeclareMathBrackets}[3]{
  \newcommand{#1}[2][]{\brk@usearg{##1}{l}{#2}##2\brk@usearg{##1}{r}{#3}}
  \brk@altname{#1}{big}{big}\brk@altname{#1}{lr}{*}}
\newcommand{\DeclareMathBiBrackets}[4]{
  \newcommand{#1}[3][]{\brk@usearg{##1}{l}{#2}##2#3##3\brk@usearg{##1}{r}{#4}}
  \brk@altname{#1}{big}{big}\brk@altname{#1}{lr}{*}}
\newcommand{\DeclareMathBiMBracketsStar}[4]{
  \newcommand{#1}[3][]{\brk@usearg{##1}{l}{#2}##2\brk@usearg{##1}{m}{#3}##3\brk@usearg{##1}{r}{#4}}
  \brk@altname{#1}{bi}{big}}
\newcommand{\DeclareMathBiBracketsStar}[4]{
  \newcommand{#1}[3][]{\brk@usearg{##1}{l}{#2}##2\brk@usearg{##1}{}{#3}##3\brk@usearg{##1}{r}{#4}}
  \brk@altname{#1}{big}{big}}

\makeatother

\DeclareMathBrackets{\brk}{(}{)}
\DeclareMathBrackets{\sbrk}{[}{]}
\DeclareMathBrackets{\set}{\{}{\}}
\DeclareMathBrackets{\abs}{|}{|}
\DeclareMathBrackets{\eval}{.}{|}
\DeclareMathBiBrackets{\comm}{[}{,}{]}
\DeclareMathBiBrackets{\acomm}{\{}{,}{\}}
\DeclareMathBrackets{\ket}{|}{\rangle}
\DeclareMathBrackets{\range}{|}{|}

\DeclareMathOperator{\Tr}{Tr}
\DeclareMathOperator{\diag}{diag}
\DeclareMathOperator{\arccosh}{arccosh}

\title{Integrable Deformations of the XXZ Spin Chain}
\author{%
Niklas Beisert\texorpdfstring{$^{a}$}{},
Lucas Fi\'evet\texorpdfstring{$^{a}$}{},
\texorpdfstring{\\}{}
Marius de Leeuw\texorpdfstring{$^{a}$}{},
Florian Loebbert\texorpdfstring{$^{b}$}{}}

\begin{document}

\pdfbookmark[1]{Title Page}{title}

\thispagestyle{empty}

{\small
\begin{center}%
\begingroup\Large\bfseries\thetitle\par\endgroup
\vspace{.5cm}

\begingroup\scshape\theauthor\par\endgroup
\vspace{5mm}%

\begingroup\itshape

$^a$
Institut f\"ur Theoretische Physik,
Eidgen\"ossische Technische Hochschule Z\"urich\\
Wolfgang-Pauli-Strasse 27, 8093 Z\"urich, Switzerland
\vspace{3mm}

$^b$
Niels Bohr International Academy,
Niels Bohr Institute,\\
Blegdamsvej 17, 2100 Copenhagen, Denmark
\vspace{3mm}

\par\endgroup
\vspace{5mm}

\begingroup\ttfamily
\texttt{%
nbeisert@ethz.ch,
lfievet@student.ethz.ch,\\
deleeuwm@phys.ethz.ch,
loebbert@nbi.dk
}
\par\endgroup

\vspace{.8cm}

\textbf{Abstract}\vspace{7mm}

\begin{minipage}{11.1cm}
We consider integrable deformations of the XXZ spin chain 
for periodic and open boundary conditions. 
In particular, we classify all long-range deformations and
study their impact on the spectrum. 
As compared to the XXX case, 
we have the z-spin at our disposal, which induces two additional 
deformations: the short-range magnetic twist and a new long-range 
momentum-dependent twist.
\end{minipage}

\end{center}
}
\vspace{.6cm}

\setcounter{tocdepth}{2}
\hrule height 0.75pt
\pdfbookmark[1]{\contentsname}{contents}
\tableofcontents
\vspace{0.8cm}
\hrule height 0.75pt
\vspace{1cm}

\setcounter{tocdepth}{2}


\section{Introduction}

Integrable spin chains constitute an important class of physical models. 
They naturally appear in the context of condensed matter physics 
and play an important \textit{r\^ole} for the AdS/CFT correspondence. 
It proves useful to distinguish between integrable nearest-neighbor (NN) 
and long-range spin chains.
The former class is characterized by a Hamiltonian density 
that acts on two neighboring spins only. 
These spin chains are rather well-understood; 
mostly in terms of the R-matrix formalism. 
Long-range spin chains on the other hand are less studied 
and seem to be more difficult to capture. 
Their most famous representatives are the Haldane-Shastry 
and Inozemtsev models \cite{Haldane:1988gg,SriramShastry:1988gh,Inozemtsev:1989yq,Inozemtsev:2002vb}. 
Long-range spin chains also appear in the context of the AdS/CFT correspondence. 
Here the dilatation operator of $\mathcal{N}=4$ super Yang--Mills (SYM) 
theory maps to an integrable spin chain Hamiltonian 
whose higher loop corrections correspond to perturbative long-range deformations 
in a small (coupling) parameter, see for instance \cite{Beisert:2010jr}.

A general framework to deal with integrable long-range spin chains 
was put forward in \cite{Bargheer:2008jt,Bargheer:2009xy} 
for periodic boundary conditions and in \cite{Loebbert:2012yd} for open chains. 
Based on a deformation equation, different generators were identified 
to explain all degrees of freedom found in a previous brute force approach 
to $\alg{gl}(N)$ symmetric long-range chains \cite{Beisert:2005wv,Beisert:2008cf}.
In this paper we extend the above results 
to a more general spin chain with two spin states (up/down) at each site. 
We require that the starting Hamiltonian is homogeneous and consists 
of NN interactions which preserve the number of up and down spins. 
The considered system is equivalent to a long-range deformed twisted version 
of the Heisenberg XXZ spin chain, and it constitutes an integrable model. 

We study general long-range deformations of the periodic Hamiltonian: 
First we explicitly compute all long-range deformations up to range four 
by the brute force method. 
Then we make contact with the deformation equation and relate the different 
degrees of freedom to various generators of long-range deformations. 
We find new types of deformation generators that can be built 
using the spin operator in z-direction. 
Among these deformations is the well known NN twist as well as 
a class of long-ranged momentum dependent twists. 
The quantum deformation parameter $\hbar$ 
does not fit into the above framework of deformations which preserves the symmetry algebra.
Then we extend the discussion to open spin chains 
where the leading deformations are computed by brute force 
and the deformations are related to their corresponding generators. 
Also here the NN boundary paramters $\xi_\LeftRight$ 
are not mapped to specific generators.
For periodic and open boundaries we classify all long-range deformations 
and their effect on the Bethe equations and the spectrum. 
The different deformations are listed in \tabref{tab:overviewclosed}
and \tabref{tab:overviewopen} in \secref{sec:conclusions}.

This paper is organized as follows. First we discuss the twisted XXZ spin chain 
with periodic boundary conditions and NN interactions. 
Subsequently we extend this to long-range spin chains 
and give the corresponding Bethe equations. 
Finally we extend the discussion to open spin chains. We end with conclusions.

\section{Periodic NN Spin Chain}
\label{section:NN}

We first consider a general closed spin chain with 
two spin states (up/down) at each site. 
We demand that the Hamiltonian is homogeneous
and consists of nearest-neighbor interactions 
which preserve the number of up and down spins.
This amounts to the so-called Heisenberg XXZ spin chain 
(with some twist parameters), 
and it constitutes an integrable model. 
We will describe its Hamiltonian 
and the corresponding spectrum through the Bethe ansatz.

\paragraph{Hamiltonian and integrability.}

The Hamiltonian on a periodic spin chain 
is given by the homogeneous sum $\oplocal{\opkern{H}}$ 
of a two-site interaction kernel $\opkern{H}$ over all 
pairs of neighboring sites $(a,a+1)$ of the chain
\[
\oplocal{\opkern{H}}:=\sum_{a=1}^L \opkern{H}_{a}.
\]
The most general two-site operator $\opkern{H}$
which conserves the number of up and down spins reads
\begin{align}\label{eqn:HNN}
\opkern{H} =\mathord{}
& h_1 (1\otimes 1) 
+ h_2 (1\otimes \sigma_z + \sigma_z\otimes 1) 
+ h_3 (\sigma_z \otimes \sigma_z ) 
\nonumber\\&
+ h_4 (\sigma_+ \otimes \sigma_-) 
+ h_5  (\sigma_- \otimes \sigma_+) 
+ h_6 (1\otimes \sigma_z - \sigma_z\otimes 1).
\end{align}
Here $\sigma_z$, $\sigma_\pm:=(\sigma_x\pm i\sigma_y)/2$ 
are the standard $2\times 2$ Pauli matrices.

Before continuing, let us look at the different terms in the Hamiltonian.
The coefficient $h_1$ of the identity operator 
determines the overall energy level,
while $h_2$ multiplies the total spin operator.
Then, $h_3$ describes the strength of the spin-spin interaction, 
and $h_4,h_5$ determine the strength of the hopping terms.
Finally, $h_6$ multiplies the structure 
$ 1\otimes \sigma_z - \sigma_z\otimes 1$
which trivially vanishes on periodic chains.

The Hamiltonian \eqref{eqn:HNN} turns out to be compatible with integrability.
Consider the following trigonometric R-matrix for the XXZ chain
\begin{align}\label{eqn:Rmatrix}
\opkern{R}(u) = &\,
e^{i\rho u}\left[
\frac{1\otimes 1 + \sigma_z\otimes\sigma_z}{2} + (\sigma_-\otimes\sigma_+)\frac{e^{i\psi u}\sinh i\hbar \kappa}{\sinh\hbar(u+i\kappa)}
+(\sigma_+\otimes\sigma_-)\frac{e^{-i\psi u}\sinh i\hbar \kappa}{\sinh\hbar(u+i\kappa)}\right.
\nonumber\\
&\left.\mathord{}+\frac{(1\otimes 1 - \sigma_z\otimes\sigma_z)}{2}\,\frac{\cosh i\phi\sinh\hbar u}{\sinh\hbar(u+i\kappa)}
+\frac{(\sigma_z\otimes1-1\otimes\sigma_z)}{2}\,\frac{\sinh i\phi\sinh\hbar u}{\sinh\hbar(u+i\kappa)}
\right].
\end{align}
Then it is readily shown that $\opkern{R}$ satisfies the Yang--Baxter equation.
In other words, this R-matrix belongs to an integrable system.

Let us pause again to discuss the parameters of $\opkern{R}$:
The coefficient $\hbar$ corresponds to the quantum deformation parameter
of the XXZ chain, whereas $\kappa$ sets a scale for the variable $u$. 
The parameter $\rho$ affects only the overall factor of $\opkern{R}$.
The coefficient $\phi$ is generated by the two-site twist  \cite{ReshTwist} 
\begin{align}
\opkern{R}_{12}&\to \opkern{F}_1 \opkern{R}_{12}\opkern{F}_2^{-1},
& \opkern{F} &= \exp \ihalf\phi\sigma_z,
\end{align}
whereas $\psi$ is generated by the one-site similarity transformation
\begin{align}\label{eqn:PsiSimTrans}
\opkern{R}(u_1-u_2) &\to (\opkern{G}(u_1)\otimes\opkern{G}(u_2))^{-1} \opkern{R}(u_1-u_2) (\opkern{G}(u_1)\otimes\opkern{G}(u_2)),
& \opkern{G}(u) &= \exp \ihalf \psi u \sigma_z.
\end{align}
Both the twist and and the similarity transformation are well-known deformations that preserve integrability. 

At the point $u=0$, the R-matrix reduces to the permutation operator.
Generically, the corresponding Hamiltonian is given by the logarithmic derivative
of the transfer matrix at this point, 
and consequently the Hamiltonian kernel $\opkern{H}$ 
is found from the R-matrix as follows
\begin{align}\label{eqn:HviaR}
\opkern{H} =  -i\lreval{\opkern{R}(u)^{-1}\frac{d}{du}\opkern{R}(u)}_{u=0}.
\end{align}
Comparing this against the explicit form of \eqref{eqn:HNN} we find
\begin{align}\label{eqn:finalcoeff}
 h_1 &= \rho +\frac{i\hbar}{2} \coth i\hbar \kappa\,, 
& h_2&=0, 
& h_3&=-\frac{i\hbar}{2}\coth i\hbar \kappa\,,
\nonumber\\
 h_4 &= \frac{-i\hbar e^{-i\phi}}{\sinh i\hbar \kappa}\,, 
& h_5 &= \frac{-i\hbar e^{i\phi}}{\sinh i\hbar \kappa}\,, 
& h_6 &= -\frac{\psi}{2}\,.
\end{align}
This relates the five free parameters $\rho,\kappa,\hbar,\psi,\phi$ of the R-matrix
to five of the degrees of freedom of the Hamiltonian 
\begin{align}
\rho &= h_1+h_3, 
& \hbar&=i\sqrt{4h_3^2-h_4h_5}, 
&\phi&=\frac{i}{2}\log\frac{h_4}{h_5}\,, 
&i\hbar \kappa &= \arccosh\frac{2h_3}{\sqrt{h_4h_5}}\,, 
&\psi &= -2h_6.
\end{align}
The remaining parameter $h_2$ is associated to the overall spin operator
which trivially commutes with $\opkern{R}$.%
\footnote{The spin operator follows from the expansion of $\opkern{R}$ around $u=\infty$
rather than around $u=0$.}
This establishes integrability of our spin chain Hamiltonian.

Hermiticity of the Hamiltonian and unitarity of the R-matrix correspond 
to $\phi,\kappa,\psi,\rho$ and $u$ being real, 
while $\hbar$ can either be real or imaginary.

It is easily seen that the model contains both the XXX and 
the conventional XXZ spin chain.
Specifying $\phi=\psi=0$ and $\kappa=1$ reduces $\opkern{R}$ 
to the conventional XXZ R-matrix,
which, in turn, can be reduced to the XXX model
by taking the limit $\hbar \to 0$.

The parameters $h_1,h_2,h_6$, or equivalently $\rho,\psi$,
do not influence the eigenvectors of the Hamiltonian
and simply shift the energy eigenvalues.
Thus, for concreteness and without loss of generality we shall set 
\begin{align}
\label{eq:simplepar}
h_1&=-h_3, &
h_2=h_6=\rho=\psi&=0.
\end{align}
Finally there is a degree of freedom corresponding to an overall rescaling of the Hamiltonian. 
This simply corresponds to a rescaling of $u$ in the R-matrix. 
In fact, by rescaling $(u,\hbar)\rightarrow (\kappa u,\hbar\kappa^{-1})$ we can set $\kappa=1$. 
Thus in what follows we will also fix the normalization of the Hamiltonian by setting $\kappa=1$.

\paragraph{Higher conserved charges.}  

Since the model is integrable,
there is a tower of commuting conserved charges 
with local kernels $\opkern{Q}_r$, $r=2,3,\ldots$,
of range $r$; 
the Hamiltonian $\opkern{H}$ is typically identified 
with the lowest charge $\opkern{Q}_2$.
It is well known that these can be generated recursively by
the so-called boost operator $\opboost{\opkern{H}}$.
To this end we must consider an infinite spin chain.
Let $\opkern{L}$ be the kernel of some local operator 
with some fixed range;
then the action of the operator $\oplocal{\opkern{L}}$ 
and of its boosted version $\opboost{\opkern{L}}$
on the infinite spin chain read
\begin{align}
\label{eq:localboostdef}
\oplocal{\opkern{L}} &:= \sum_{a} \opkern{L}_a, 
&
\opboost{\opkern{L}} &:= \sum_{a} a  \opkern{L}_a.
\end{align}
The kernels $\opkern{Q}_r$ of the higher charges
are now generated recursively 
by the boosted Hamiltonian \cite{Tetelman:1981xx}
\begin{align}\label{highercharges}
\oplocal{\opkern{Q}_{r}}  = \frac{i}{r-1} \comm[big]{\opboost{\opkern{Q}_2}}{\oplocal{\opkern{Q}_{r-1}}}.
\end{align}
Here we identify the lowest charge $\opkern{Q}_2$ 
with the Hamiltonian $\opkern{H}$ fixing some parameters
as explained above in \eqref{eq:simplepar}
\begin{align}
\opkern{Q}_2 &=
  h_3 (\sigma_z \otimes \sigma_z- 1\otimes 1) 
+ h_4 (\sigma_+ \otimes \sigma_-) 
+ h_5 (\sigma_- \otimes \sigma_+) 
.
\end{align}
Note that the commutator of a boosted charge with another charge
is again a well-defined local operator.

Let us explicitly spell out the charges $\opkern{Q}_{3,4}$ since they will play a
role in our discussion of long-range deformations.
The range-three operator is readily worked out from \eqref{highercharges} 
\begin{align}\label{eqn:Q3NN}
\opkern{Q}_3  
&= \frac{i\hbar^2}{2}\,\frac{\coth i\hbar \kappaR}{\sinh i\hbar \kappaR}
\left[\frac{e^{2i\phi}}{\cosh i\hbar \kappaR}\, (\sigma_-\otimes\sigma_z\otimes\sigma_+) -
e^{i\phi}\, (\sigma_-\otimes\sigma_+\otimes\sigma_z + \sigma_z\otimes\sigma_-\otimes\sigma_+)\right.
\nonumber\\
&\qquad\left.+e^{-i\phi}\, (\sigma_+\otimes\sigma_-\otimes\sigma_z +  \sigma_z\otimes\sigma_+\otimes\sigma_-)
- \frac{e^{-2i\phi}}{\cosh i\hbar \kappaR}\, (\sigma_+\otimes\sigma_z\otimes\sigma_-)
\right].
\end{align}
We furthermore find
\begin{align}
\label{eq:Q4NN}
\opkern{Q}_4  =\mathord{}
& \frac{i\hbar^3}{3\tanh^3 i\hbar \kappaR}\left[
\frac{1+\cosh^2 i\hbar \kappaR}{\cosh^3 i\hbar \kappaR}
(e^{i\phi} \sigma_-\otimes\sigma_+ +e^{-i\phi} \sigma_+\otimes\sigma_-)  + 
\frac{ \sigma_z\otimes\sigma_z}{\cosh^2 i\hbar \kappaR} 
\right.
\nonumber\\
& +
\frac{1}{\cosh i\hbar \kappaR}(2e^{2i\phi} \sigma_-\otimes 1 \otimes\sigma_+ 
+ 2e^{-2i\phi} \sigma_+\otimes 1 \otimes\sigma_- + \sigma_z\otimes 1\otimes \sigma_z)  \nonumber\\
& -\frac{1}{\cosh^3 i\hbar \kappaR}( e^{3i\phi}\sigma_-\otimes\sigma_z\otimes\sigma_z\otimes\sigma_+ +
e^{-3i\phi}\sigma_+\otimes\sigma_z\otimes\sigma_z\otimes\sigma_- )\nonumber\\
& -\frac{1}{\cosh i\hbar \kappaR}( e^{i\phi}\sigma_z\otimes\sigma_-\otimes\sigma_+\otimes\sigma_z +
e^{-i\phi}\sigma_z\otimes\sigma_+\otimes\sigma_-\otimes\sigma_z )\nonumber\\
&+ \frac{e^{2i\phi}}{\cosh^2i\hbar \kappaR}( \sigma_-\otimes\sigma_z\otimes\sigma_+\otimes\sigma_z +
\sigma_z\otimes\sigma_+\otimes\sigma_z\otimes\sigma_-
-2\sigma_-\otimes\sigma_+\otimes\sigma_-\otimes\sigma_+)\nonumber\\
&+ \frac{e^{-2i\phi}}{\cosh^2 i\hbar \kappaR}( \sigma_+\otimes\sigma_z\otimes\sigma_-\otimes\sigma_z +
\sigma_z\otimes\sigma_-\otimes\sigma_z\otimes\sigma_+
-2\sigma_+\otimes\sigma_-\otimes\sigma_+\otimes\sigma_-)\nonumber\\
&+ \left.\frac{2}{\cosh^2 i\hbar \kappaR}(\sigma_+\otimes\sigma_-\otimes\sigma_-\otimes\sigma_+ + 
\sigma_-\otimes\sigma_+\otimes\sigma_+\otimes\sigma_-)\right].
\end{align}
Notice that the above operators are trivially Hermitian provided that $\phi$ 
is real and $\hbar\kappaR$ is either real or imaginary.

\paragraph{Equivalence classes for local operators.}  

The above construction using the boost operator
relies on an identification of operator kernels
that will be particularly relevant to the
construction of long-range interactions. 
Let us therefore discuss it in some more detail.

On closed and on infinite spin chains, 
a local operator kernel $\opkern{L}$ is
equivalent to the kernel with an 
additional leg attached to the right or to the left
which acts as the identity%
\footnote{For a closed chain this statement holds
only if the range of $\opkern{L}$ is less than the length of the chain. 
Otherwise the action of 
$\opkern{L}\otimes 1$ and $1\otimes\opkern{L}$ has no proper meaning.
Therefore we can only identify 
local operators on infinite chains 
with local operators on closed chains
as long as the range is smaller than the length of the chain.
This ambiguity is the so-called wrapping problem.}
\[
\opkern{L}\otimes 1 \simeq \opkern{L} \simeq 1\otimes\opkern{L}.
\]
This equivalence originates from the identity 
of the associated homogeneous local operators 
on periodic chains
\[\label{eq:L1=1L=L}
\oplocal{\opkern{L}\otimes 1}=\oplocal{\opkern{L}}=\oplocal{1\otimes\opkern{L}}.
\]

The boost construction \eqref{highercharges}
therefore does not uniquely specify the charge kernels
$\opkern{Q}_{r}$, but merely their equivalence class 
$\oplocal{\opkern{Q}_{r}}$. Alternatively, we could write 
the boost construction as 
\begin{align}
\opkern{Q}_{r} \simeq \frac{i}{r-1} \comm[big]{\opboost{\opkern{Q}_2}}{\opkern{Q}_{r-1}}.
\end{align}
One can convince oneself that the choice of representative of
the equivalence classes does not play a role for the
construction of the higher charges.%
\footnote{In fact, it does play a role in the
definition of $\opboost{\opkern{Q}_2}$, 
but only up to a contribution of the local operators
$\oplocal{1}$ and $\oplocal{\sigma_z}$ which commute
with all charges $\opkern{Q}_r$.}

In that regard it makes perfect sense to 
work with linear combinations of local operators of different length,
see \textit{e.g.}\ \eqref{eq:Q4NN}.
For the definition of the (boost) action on a chain in \eqref{eq:localboostdef},
one should think of $\opkern{L}_a$
as an operator whose first leg acts on site $a$ of the spin chain.

\paragraph{Bethe ansatz.} 
The eigenvalues of the conserved charges, including the Hamiltonian, 
can be found by applying the algebraic Bethe ansatz \cite{Faddeev:1996iy}. 
We introduce the Lax operator $\opkern{L}_{a,n}$ 
and the corresponding transfer matrix  $\oper{T}$ as
\begin{align}
&\opkern{L}_{a,n} = \opkern{R}_{n,a}(u-\ihalf\kappaR),
&& \oper{T}(u) =\Tr_a \opkern{L}_{a,1} \ldots  \opkern{L}_{a,L}.\label{eq:transfer}
\end{align}
The eigenvalues $T$ of the transfer matrix can be found by following \cite{Faddeev:1996iy}:
\begin{align}
T(u| \{ u_i \}) =&
\prod_{i=1}^M e^{-i \phi}\frac{\sinh\hbar(u-u_i-i\kappaR)}{\sinh\hbar(u-u_i)}
+\left[e^{i\phi}\frac{\sinh\hbar (u-\ihalf\kappaR) }{\sinh\hbar(u+\ihalf\kappaR)}\right]^L
\prod_{i=1}^M  e^{-i \phi}
\frac{\sinh\hbar(u-u_i+i\kappaR)}{\sinh\hbar(u-u_i)}
\,,
\end{align}
where this eigenvalue corresponds to a state with $M$ magnons with rapidities $u_1,\ldots u_M$. 
Cancellation of the superficial poles at $u=u_i$ results in the Bethe equations
\begin{align}\label{eqn:BAE}
\left[e^{-i \phi}
\frac{\sinh\hbar(u_k+\ihalf\kappaR) }{\sinh\hbar(u_k-\ihalf\kappaR)}\right]^L =
\mathop{\prod_{i=1}^M}_{i\neq k} 
 \frac{\sinh\hbar(u_k-u_i+i\kappaR) }{\sinh\hbar(u_k-u_i-i\kappaR)}.
\end{align}
The magnon momentum and energy are then related to the first two terms
in the expansion of the transfer matrix.
More precisely, the momentum $P$ 
\footnote{The momentum $P$ determines the eigenvalue $e^{iP}$ 
of the cyclic shift operator by one site.
Consequently, $P$ is defined only modulo $2\pi$.}
and energy $E$ are obtained by the logarithm and logarithmic derivative at the point $u=\ihalf\kappaR$,
respectively
\begin{align}
P&=\sum_{k=1}^M p(u_k),
&
E&=Q_2=\sum_{k=1}^M q_2(u_k),
\end{align}
where the magnon charges are given by 
\begin{align}\label{eqn:pandE}
p(u) &=-\phi + \frac{1}{i}\log\frac{\sinh\hbar(u+\ihalf\kappaR)}{\sinh\hbar(u-\ihalf\kappaR)}\,, 
&
q_2(u) &=\frac{i\hbar}{\tanh\hbar(u+\ihalf\kappaR)}-\frac{i\hbar}{\tanh\hbar(u-\ihalf\kappaR)}\,.
\end{align}
Notice that $dp/du = - q_2$. 
The transfer matrix actually generates the complete tower 
of conserved charges \eqref{highercharges} by a logarithmic expansion
in the rapidity $u$. 
The eigenvalues $Q_r$ of the higher conserved charges
can also be written as a sum over one-magnon eigenvalues $Q_r = \sum_k q_r(u_k)$. 
One can show that the one-magnon eigenvalues $q_r$ satisfy the following recursion relation
\begin{align}\label{eqn:recursiveq}
q_r = \frac{1}{1-r}\, q_2 \frac{d}{d p}q_{r-1}(p)  
= \frac{1}{r-1}\, \frac{d}{d u}q_{r-1}(u).
\end{align}
This completely determines the spectrum of the described NN spin chain. 

Let us briefly show that the conserved charges generated 
by the transfer matrix indeed give rise to the correct eigenvalues 
of the explicit Hamiltonian \eqref{eqn:HviaR} and $\opkern{Q}_3$ \eqref{eqn:Q3NN} 
by considering one magnon states. 
Define the vacuum state as the state where all spins are pointing up. 
The one-magnon state is then defined as
\begin{align}
&\ket{p} = \sum_{n=1}^Le^{ipn}(\sigma_-)_n \ket{0}, 
&&
\ket{0} = \ket{\mathord{\uparrow}\ldots\mathord{\uparrow}}.
\end{align}
It is easily seen that $\ket{p} $ is an eigenstate of $\opkern{Q}_2$ and $\opkern{Q}_3$, 
provided that $e^{ipL}=1$, with eigenvalues
\begin{align}
&q_2(p) = 2i\hbar \,\frac{\cosh i\hbar - \cosh i(p+\phi)}{\sinh i\hbar }\,,
&& q_3(p) = -\hbar \,\frac{\sinh i(p+\phi)}{\sinh i\hbar }\,q_2(p).
\end{align}
These eigenvalues indeed coincide with \eqref{eqn:pandE} and \eqref{eqn:recursiveq}.


\section{Long-Range Deformations}
\label{section:LRpert}

Let us now discuss the possible perturbative long-range deformations
of our XXZ spin chain. We will denote the range of an operator $\opkern{L}$ by $\range{\opkern{L}}$.
An operator $\opkern{L}$ of range $\range{\opkern{L}}=r$ 
acts on at most $r$ spin chain sites at the same time. 
In what follows we append an index NN to several nearest-neighbor quantities 
computed in the previous section in order to distinguish 
them from their long-range counterparts considered here.

\paragraph{Formalism.}
Any NN integrable spin chain has a tower of independent conserved charges 
$\{\opkern{Q}_r\}_{r=2,3,\ldots}$,%
\footnote{Sometimes $\opkern{Q}_1$ 
is regarded as the momentum and is also a conserved charge.}
such that $\opkern{Q}_{2} = \opkern{H}$ and $\range{\opkern{Q}_r} = r$. Such a spin chain can be perturbatively 
extended to a long-range spin chain, meaning that all the conserved charges 
(including the Hamiltonian) have infinite range. The deformation is based 
on the introduction of a parameter $\lambda$ in which all the charges can be expanded
\begin{align}
\opkern{Q}_r(\lambda) = \sum_{k=0}^{\infty} \opkern{Q}^{(k)}_r \lambda^k,
\end{align}
such that $\range{\opkern{Q}^{(k)}_r} = r+k$ and $\opkern{Q}_r(0)=\opkern{Q}_r^{(0)} = \opkern{Q}_r$.
This new family $\{\opkern{Q}_r(\lambda)\}$ then defines an integrable model provided that
\begin{align}\label{eqn:pertcommrel}
&\comm[big]{\oplocal{\opkern{Q}_r(\lambda)}}{\oplocal{\opkern{Q}_s(\lambda)}} 
=0, &\Leftrightarrow& & 
\sum_{n=0}^{k} \comm[big]{\oplocal{\opkern{Q}_r^{(n)}}}{\oplocal{\opkern{Q}_s^{(k-n)}}} = 0
\quad \forall k.
\end{align}
For $k=0$ this trivially reduces to the commutation relations of the
nearest-neighbor spin chain we started with.
At this point we would also like to point out the trivial perturbative
deformations for which $\range{\opkern{Q}^{(k)}_r}$ is not increasing when $k\to k+1$.
They are simply obtained by making the coefficients of the Hamiltonian functions of $\lambda$.
Note that for a given spin chain this definition of long-range charges 
is only valid up to the (wrapping) order where the operator wraps the whole state. 

\paragraph{First order.} Let us, by explicit computation, 
find the different long-range deformations of the 
XXZ
spin chain perturbatively.
The first non-trivial commutation relation from \eqref{eqn:pertcommrel}
that we want to be satisfied is
\begin{align}\label{eqn:longrangepert}
\comm[big]{\oplocal{\opkern{Q}_2^{(1)}}}{\oplocal{\opkern{Q}_3^{(0)}}} +
\comm[big]{\oplocal{\opkern{Q}_2^{(0)}}}{\oplocal{\opkern{Q}_3^{(1)}}} = 0,
\end{align}
where again $\oplocal{\opkern{L}}$ represents the homogeneous sum 
of the local kernel $\opkern{L}$.
We would like to compute all non-trivial solutions to the above equation
that give rise to a consistent long-range spin chain.

To this end, it is useful to divide the above equation into a homogeneous
and a non-homogeneous part. Solutions $\bar{\opkern{Q}}_2^{(1)},\bar{\opkern{Q}}_3^{(1)}$ 
to the homogeneous part are defined such that
\begin{align}
\comm[big]{\oplocal{\bar {\opkern{Q}}_2^{(1)}}}{\oplocal{\opkern{Q}_3^{(0)}}} =
\comm[big]{\oplocal{\opkern{Q}_2^{(0)}}}{\oplocal{\bar{\opkern{Q}}_3^{(1)}}} = 0.
\end{align}
The explicit solutions for $\bar{\opkern{Q}}_2^{(1)},\bar{\opkern{Q}}_3^{(1)}$ are
\begin{align}\label{eqn:trivdeform}
\bar{\opkern{Q}}_2^{(1)}&= a_1 +  a_2\,\opkern{Z}
+ a_3\,\opkern{Q}_2^{(0)} + a_4\,\opkern{Q}_3^{(0)},
\\
\bar{\opkern{Q}}_3^{(1)}&= b_1 +  b_2\,\opkern{Z}
+ b_3\,\opkern{Q}_2^{(0)} + b_4\,\opkern{Q}_3^{(0)} + b_5\,\opkern{Q}_4^{(0)},
\end{align}
where $a_i,b_i$ are free parameters.
Here we have introduced $\opkern{Z}$ as  
a combination of the unit $1$ and spin operator $\sigma_z$ 
\[\label{eq:zdef}
\opkern{Z} = \half(1-\sigma_z).
\]
It is particularly convenient 
for the Bethe ansatz because
it counts the number of spin flips $M$ above the vacuum
in the magnon excitation picture.
The above long-range deformations are trivial in the sense that they either
correspond to adding the total spin or identity operator to our Hamiltonian
or simply mix the already known conserved charges.
Consequently, the effect on the spectrum is known and given by
\begin{align}
\bar Q_2=  Q_2^{(0)} + \lambda( a_1 L + a_2 M + a_3 Q^{(0)}_2 + a_4 Q^{(0)}_3)
 +\ldots.
\end{align}
This then leaves us with the space of solutions of \eqref{eqn:longrangepert}
that do not commute independently.
We find six possible deformations of the Hamiltonian satisfying \eqref{eqn:longrangepert}
\begin{align}\label{eqn:firstorder}
&a_5 (e^{-i\phi}\sigma_+\otimes\sigma_- + e^{i\phi}\sigma_-\otimes\sigma_+)+\nonumber\\
&ia_6 (e^{-i\phi}\sigma_+\otimes\sigma_- -
e^{i\phi}\sigma_-\otimes\sigma_+)+\nonumber\\
&a_7 (\sigma_z\otimes1\otimes\sigma_z +
2e^{2i\phi}\sigma_-\otimes1\otimes\sigma_+ +2e^{-2i\phi}\sigma_+\otimes1\otimes\sigma_-) + \nonumber\\
&ia_8(e^{2i\phi}\sigma_-\otimes1\otimes\sigma_+ -e^{-2i\phi}\sigma_+\otimes1\otimes\sigma_-)+\nonumber\\
&a_9[(\sigma_-\otimes\sigma_+ + \sigma_+\otimes\sigma_-)\otimes\sigma_z-
\sigma_z\otimes(\sigma_-\otimes\sigma_+ + \sigma_+\otimes\sigma_-)]+\nonumber\\
&ia_{10}[(\sigma_-\otimes\sigma_+ -\sigma_+\otimes\sigma_-)\otimes\sigma_z+
\sigma_z\otimes(\sigma_-\otimes\sigma_+ - \sigma_+\otimes\sigma_-)].
\end{align}
It is easily seen that our deformations are Hermitian if and only if the
coefficients $a_i$ are real.
To each deformation $\opkern{Q}_2^{(1)}$ of the Hamiltonian corresponds a deformation $\opkern{Q}_3^{(1)}$ 
of the next higher charge that involves the same set of parameters.
The deformations proportional to $a_1,a_3,a_5,a_6$ 
simply arise from making the various parameters in the Hamiltonian $\lambda$-dependent.
Such a deformation does not affect the range of the Hamiltonian 
and the higher charges and its effect on the spectrum is once again simple. 
It is actually a deformation of the NN model rather than a long-range deformation. 

This means that at this order, there are exactly four non-trivial long-range deformations 
of the spin chain corresponding to non-trivial values of $a_{7,8,9,10}$. 
Their effect on the spectrum of the spin chain will be discussed in the next section.

Computing the commutator between $\opkern{Q}_2,\opkern{Q}_4$ to first order in $\lambda$ 
shows a similar structure. To every non-trivial long-range deformation of the Hamiltonian 
corresponds a unique \emph{non-trivial} deformation of central charge $\opkern{Q}_4$. 
Assuming that this structure does not break down for the other conserved charges, 
we see that we have found all long-range deformations that preserve integrability 
to leading order in $\lambda$.

\paragraph{Second order.} The next-to-leading order in $\lambda$ relates $\opkern{Q}^{(2)}_r$ 
to the leading order deformations studied in the previous paragraph. 
To second order in $\lambda$, the commutator between the Hamiltonian 
and the next central charge becomes
\begin{align}
\comm[big]{\oplocal{\opkern{Q}_2^{(2)}}}{\oplocal{\opkern{Q}_3^{(0)}}} +
\comm[big]{\oplocal{\opkern{Q}_2^{(1)}}}{\oplocal{\opkern{Q}_3^{(1)}}} +
\comm[big]{\oplocal{\opkern{Q}_2^{(0)}}}{\oplocal{\opkern{Q}_3^{(2)}}} =0.
\end{align}
Solving this gives 23 possible solutions for $\opkern{Q}^{(2)}_2$. 
The solutions clearly depend on $\opkern{Q}_2^{(1)}$, hence deforming the Hamiltonian 
with a term linear in $\lambda$ only gives rise to an integrable spin chain 
if the Hamiltonian also gets a corresponding correction at order $\lambda^2$. 
This pattern will continue and thus deforming our Hamiltonian 
will automatically imply that all operators become of infinite range. 
However, taking $\lambda$ to be small we can simply work to a given order in $\lambda$ 
and actually avoid this problem. 
At $\lambda^2$ we find 23 degrees of freedom in addition to the 
10 that we found previously at order $\lambda$.


\section{Deformation Equation}

In the previous section we identified the first few non-trivial long-range deformations
of the Hamiltonian that preserve integrability.
In this section we will describe them in a constructive way by means of the
so-called deformation equation.
This allows us to classify all the deformations at any given order.

\subsection{Deformation Equation} 

Consider the first order differential equation
\begin{align}\label{eqn:DefEQN}
\frac{d}{d \lambda} \oplocal{\opkern{Q}_r(\lambda)} &= 
i \comm[big]{\oper{X}(\lambda)}{\oplocal{\opkern{Q}_r(\lambda)}}, 
&
 \opkern{Q}_r(0) &= \opkern{Q}_r.
\end{align}
It is readily seen that the solutions $\opkern{Q}_r(\lambda)$ of \eqref{eqn:DefEQN}
still commute and hence, provided that these operators are homogeneous and local,
they give rise to an integrable spin chain.
Explicitly, one can recursively solve the higher charges as follows
\begin{align}
\oplocal{\opkern{Q}^{(k+1)}_r} &= 
\frac{i}{k+1}\sum_{i=0}^{k}\comm[big]{\oper{X}^{(i)}}{\oplocal{\opkern{Q}^{(k-i)}_r}},
&
\oper{X} &= \sum_{i=0}^\infty \oper{X}^{(i)} \lambda^i.
\end{align}
There are three types of operators $\oper{X}$ 
such that the solutions $\opkern{Q}_r$ of \eqref{eqn:DefEQN} are local operators:
\begin{itemize}
 \item local operators $\oplocal{\opkern{L}}$,
 \item boosted charges $\opboost{\opkern{Q}_r}$ and $\opboost{\opkern{Z}}$,
 \item bilocal charges $\opbilocal{\opkern{Q}_r}{\opkern{Q}_s}$ and $\opbilocal{\opkern{Z}}{\opkern{Q}_r}$. 
\end{itemize}
The boosted operator can be any operator that commutes with the Hamiltonian. 
Obviously, these are the conserved charges $\opkern{Q}_r$, 
but also $1,\sigma_z$. 
Since $1$ yields a trivial deformation, 
we choose to add the operator $\opkern{Z} = \half(1-\sigma_z)$
defined in \eqref{eq:zdef}
which has a vanishing vacuum eigenvalue.
This will be convenient for the physical interpretation 
of corresponding deformations in the Bethe equations.

Finally, for any commuting family of charges $\{\opkern{Q}_r(\lambda)\}$ 
the linear combinations $\opkern{Q}_r(\lambda)\rightarrow \gamma_{r,s}\opkern{Q}_s(\lambda)$ 
trivially give rise to an acceptable deformation. 
Notice that when $r=s+1$ we could write it in the form of the deformation equation 
since $\opkern{Q}_{r+1} \simeq \frac{i}{r-1}[\opboost{\opkern{Q}_2},\opkern{Q}_r]$. 
We will now discuss in detail the different types of deformations.

\paragraph{Local operators.} 

For any local operator $\oper{X}=\oplocal{\opkern{L}}$, 
the commutator $\comm{\oper{X}}{\opkern{Q}_r}$ is trivially local again.
Let $\opkern{L}$ be an operator of range $\range{\opkern{L}}=l$, 
then the range of the deformed operator
is simply $\comm{\oplocal{\opkern{L}}}{\opkern{Q}_r} = r+l-1$.
In this case the deformation equation \eqref{eqn:DefEQN} 
is solved exactly by 
$\opkern{Q}_{r}(\lambda) =  e^{i\lambda \oper{X}} \opkern{Q}_r e^{-i\lambda \oper{X}}$,
which is simply a globally defined similarity transformation on the closed spin chain. 
As such, these deformations do not affect the spectrum of the conserved charges $\opkern{Q}_r$.
Notice that $\opkern{L}=\opkern{Q}_s$ induces trivial transformations, 
and has no associated deformation parameters.

\paragraph{Boosted charges.}

As discussed in \secref{section:NN}, the commutator of a boosted charge
with the conserved charges results again in a local operator.
In particular, commutators with the boosted Hamiltonian generate
all higher charges for the NN model.
In general, $\range{\comm{\opboost{\opkern{Q}_r}}{\opkern{Q}_s}} = r+s-1$,
however it can be shown that \cite{Bargheer:2008jt}
\begin{align}\label{eqn:ReduceRankBoost}
\comm[big]{\opboost{\opkern{Q}_r}}{\opkern{Q}_s} 
\simeq \opkern{Q}_{r+s-1} + \text{lower range}.
\end{align}
In other words, if we combine a basis transformation 
with a boost we can reduce the length by at least one.
This opens up one additional deformation of the Hamiltonian at range $k$. 

Furthermore we can deform with the boost $\opboost{\opkern{Z}}$
of the spin operator $\opkern{Z}$ 
which does not increase the range of the deformed charge. 
Hence, this is a NN deformation which generates the twist $\phi$ 
that was already introduced in \secref{section:NN}.

\paragraph{Bilocal charges.}

The last class of operators
that generate acceptable long-range deformations through the deformation equation
are the bilocal charge. Given two local operators $\opkern{L}_r,\opkern{L}_s$, 
we define 
the corresponding bilocal operator $\opbilocal{\opkern{L}_r}{\opkern{L}_s}$ as
\begin{align}\label{eq:defbiloc}
\opbilocal{\opkern{L}_r}{\opkern{L}_s} 
= \half\sum_{a,k}\Theta(k-\half(\range{\opkern{L}_r}-\range{\opkern{L}_s}))
\acomm{\opkern{L}_r(a)}{\opkern{L}_s(a+k)},
\end{align}
where $\acomm{\cdot}{\cdot}$ is the usual anti-commutator and $\Theta$ 
is the step function $\Theta(x)=0,\half,1$ for $x<0,x=0,x>0$, respectively. 
Local deformations are then obtained by inserting the local charges into the bilocal operator, 
\textit{e.g.}\ by deforming with $\opbilocal{\opkern{Q}_r}{\opkern{Q}_s}$ or $\opbilocal{\opkern{Z}}{\opkern{Q}_r}$.

\subsection{Finding the Deformations}

Let us now reproduce the deformations that we found by explicit computations
in the previous section.

\paragraph{First orders.}

The deformations \eqref{eqn:trivdeform} with coefficients $a_{1,2,3,4}$ 
are simply obtained by a change of the basis of charges. 
Furthermore, the deformations with parameter $a_{5,6}$ 
are obtained by making the coefficients of the Hamiltonian $\lambda$-dependent. 
However, it is instructive to notice that the deformation $a_6$ 
can actually be generated by the deformation equation 
for $\oper{X}=\opboost{\opkern{Z}} $, 
whereas the $a_5$ contribution can not.

In the end, this leaves us with the non-trivial deformations $a_{7},\ldots a_{10}$. 
A quick counting reveals that we can get a non-trivial 
result from the deformation equation at this order from two local operators, 
one bi-local operator and a boost. In other words, the degrees of freedom match.

Let us first study the deformations coming form local operators. 
These need to be of range two and spin preserving,
which leads us to
\begin{align}
\opkern{L} = a_9(\sigma_+\otimes\sigma_- - \sigma_+\otimes\sigma_-) 
+ ia_{10}(\sigma_+\otimes\sigma_- + \sigma_+\otimes\sigma_-).
\end{align}
Perturbatively solving the deformation equation produces exactly
the correct long-range deformations from \eqref{eqn:firstorder}.

Next we turn to the bilocal operator. 
Obviously, there is only one such operator that generates a deformation 
of the Hamiltonian of range three, namely $\opbilocal{\opkern{Z}}{\opkern{Q}_2^{(0)}}$. 
The commutator with the Hamiltonian is readily computed to be the deformation 
corresponding to $a_8$ from \eqref{eqn:firstorder} combined with some NN deformations.

Finally, there is only the boost left.
In order to get the correct range we have to combine it with a basis transformation
and subtract $\opkern{Q}^{(0)}_4$, resulting in
\begin{align}
\opkern{Q}^{(1)}_2 =\mathord{}& \bigcomm{\opboost{\opkern{Q}^{(0)}_3}}{\opkern{Q}^{(0)}_2} 
- \sfrac{3}{2}  i \opkern{Q}^{(0)}_4\nonumber\\
=\mathord{}& 
6ih_3(-2h_4^2\,\sigma_+\otimes1\otimes\sigma_-  - h_4h_5  \sigma_z\otimes1\otimes\sigma_z  
-2h_5^2 \sigma_-\otimes1\otimes\sigma_+ )+ \text{NN}.
\end{align}
We find the final remaining degree of freedom $a_7$ from \eqref{eqn:firstorder} 
apart from the range two terms, whose exact form is unimportant since they are 
NN deformations and can again be absorbed into $a_{5,6}$.

A quick computation shows that at the next order we also find all 23 possible 
deformations by similar arguments. 
For completeness, let us briefly enumerate the 13 new operators 
that generate these deformations. 
There is one new boost $\opboost{\opkern{Q}_4}$, 
one new basis transformation $\opkern{Q}_4$, 
two new bilocal operators $\opbilocal{\opkern{Z}}{\opkern{Q}_3^{(0)}},\opbilocal{\opkern{Q}_2^{(0)}}{\opkern{Q}_3^{(0)}}$ 
and finally 9 local operators:
\begin{align}
&  \sigma_+ \otimes 1 \otimes \sigma_-,
&& \sigma_- \otimes 1 \otimes \sigma_+,
&&\sigma_z \otimes \sigma_z \otimes \sigma_z,\nonumber\\
&  \sigma_+ \otimes \sigma_z \otimes \sigma_-,
&& \sigma_- \otimes \sigma_z \otimes \sigma_+,
&& \sigma_+\otimes \sigma_- \otimes \sigma_z,\\
&  \sigma_z \otimes \sigma_+ \otimes \sigma_-,
&& \sigma_z \otimes \sigma_- \otimes \sigma_+,
&& \sigma_z \otimes 1 \otimes \sigma_z.\nonumber
\end{align}
Notice that \textit{a priori} there is one more non-trivial 
local operator $\sigma_-\otimes \sigma_+ \otimes \sigma_z$ 
but since it can be expressed by $\opkern{Q}_3$ and the other local operators 
it does not generate an independent deformation. 

\paragraph{Higher orders.} 

We can enumerate the different possible deformations 
of the spin chain Hamiltonian explicitly. 
A deformation such that the deformed Hamiltonian will be of range $\leq k$ 
is simply a combination of the various deformations outlined in this section.

Firstly, there are the basis transformations that are of the form
\begin{align}
\opkern{Q}_2 \rightarrow a_0 + a_1 \opkern{Z} + a_2 \opkern{Q}_2 + \ldots a_k \opkern{Q}_k.
\end{align}
This obviously provides $N_{\text{basis}}=k+1$ independent deformations.

Next, there are the deformations obtained 
by using local operators in the deformation equation. 
Since the commutator of a local operator of range $r$ 
and the Hamiltonian is a local operator of range $r+1$ 
we need to enumerate local operators of range $\leq k-1$. 
Local operators that 
preserve the total spins up and down
can be classified easily.
In particular, for a range $\leq r$ operator the number of independent ways 
to map a state with $l$ spins down to some state with $l$ spins down 
is simply $\fracchoose{r}{l}^2$. 
Therefore, the total number of local operators is given by
\[
\fracchoose{2r}{r}=
\sum_{l=0}^r \fracchoose{r}{l}^2.
\]
However, this counting of course includes trivial operators 
that vanish on closed spin chains. 
In other words, we have to identify operators according to \eqref{eq:L1=1L=L}. 
The number of vanishing operators is 
\[
\fracchoose{2r-2}{r-1}-1.
\]
Because $\oplocal{\opkern{Q}_2}$ commutes with $\oplocal{1},\oplocal{\opkern{Z}},\oplocal{\opkern{Q}_r}$ 
these do not give rise to non-trivial deformations. 
Summarizing, we are left with
\begin{align}
N_{\text{local}} = \fracchoose{2k-2}{k-1} - \fracchoose{2k-4}{k-2} - k +1
\end{align}
local operators that generate non-trivial deformations of range at most $k$.

Furthermore, we have the deformations generated by boosts
\begin{align}
\phi\, \opboost{\opkern{Z}} + \alpha_2 \opboost{\opkern{Q}_2} 
+ \alpha_3 \opboost{\opkern{Q}_3} + \ldots + \alpha_{k} \opboost{\opkern{Q}_{k}},
\end{align}
where we tacitly assume that we apply the length reducing procedure outlined 
in \eqref{eqn:ReduceRankBoost}. Now since $\opboost{\opkern{Q}_2}$ simply generates $\opkern{Q}_3$, 
which was already accounted for as a basis transformation, 
we obtain $N_{\text{boost}} = k-1$ different boost deformations. 
Recall that $ \opboost{\opkern{Z}}$ is actually a NN deformation 
since it does not increase the range of the Hamiltonian but rather 
makes the parameter $\phi$ dependent on the coupling $\lambda$. 

This finally leaves us with bilocal operators that are built from 
two conserved quantities. Bilocal operators $\opbilocal{\opkern{Q}_r}{\opkern{Q}_s}$ 
can always be chosen 
such that $r<s$. 
The commutator $[\opbilocal{\opkern{Q}_r}{\opkern{Q}_s},\opkern{Q}_2]$ 
is then an operator of range $s+1$.%
\footnote{Note that this is only true for a specific choice 
of local regularization of the bilocal operator \eqref{eq:defbiloc} 
which corresponds to a similarity transformation in the deformation equation.}
Hence, we find
\begin{align}
N_{\text{bilocal}} =\half(k-1)(k-2).
\end{align}

There are two permissible NN deformations that are still unaccounted for, 
namely $\hbar\rightarrow\hbar(\lambda)$ and $\psi\rightarrow\psi(\lambda)$. 
In fact, the latter cannot appear since 
it multiplies a structure that vanishes on closed chains.
As far as we can tell, 
the relevant deformation $\hbar\rightarrow\hbar(\lambda)$ 
cannot be obtained via the formalism 
of the deformation equation.
Actually, this deformation would allow us to deform the 
XXX spin chain to the XXZ spin chain.
As such, $\hbar$ is a parameter of the 
quantum affine group underlying the XXZ model. 
Since the deformation equation preserves the algebra by construction, 
it evidently cannot deform the parameter $\hbar$.

Thus the total number of deformations of the Hamiltonian 
that have range at most $k$ reads
\begin{align}
N_{\text{tot}} = \fracchoose{2k-2}{k-1} - \fracchoose{2k-4}{k-2} + \half(k^2-k+6).
\end{align}
Evaluating this for $k=3,4$ we find $10,23$ which perfectly 
agrees with deformations we identified in \secref{section:LRpert}.


\section{Asymptotic Spectrum}

We study how the long-range deformations affect the Bethe equations
for a closed chain. We will consider rotations on the basis
of charges only as far as they are required to minimize
the range of interactions. We also assume that the charge operators 
have a vanishing vacuum eigenvalue.

\paragraph{Deformed Bethe Equations.}

In \cite{Bargheer:2009xy, Bargheer:2008jt}
it was shown that 
the long-range Bethe equations for $M$ magnons
on a closed chain of length $L$ take the generic form
(up to the adjustment of the scattering factor for XXZ chains)
\[\label{eqn:longrangeBAE}
\exp\brk[big]{ip(u_k)L}=\mathop{\prod_{j=1}^M}_{j\neq k} 
\exp\brk[big]{-2i\theta(u_k,u_j)}
\frac{\sinh\hbar(u_k-u_j+i\kappaR)}
{\sinh\hbar(u_k-u_j-i\kappaR)}
\,.
\]
There are three types of deformations compared to the 
NN chain at leading order:
The magnon momentum $p(u_k)$ as a function of the magnon rapidity $u_k$ 
receives corrections in $\lambda$.
The q-deformation parameter $\hbar$ 
in the NN scattering factor
becomes a function $\hbar(\lambda)$ of the coupling constant $\lambda$.
There are additional contributions to the phase
summarized in the function $\theta(u_k,u_j)$.

Let us first discuss the phase $\theta(u_k,u_j)$
associated to bilocal deformations. 
In our framework it takes the form
\[\label{eqn:dressphase}
\theta(u_k,u_j)=
\sum_{s>r=2}^\infty \beta_{r,s}\brk[big]{q_r(u_k)q_s(u_j)-q_s(u_k)q_r(u_j)}
+
\sum_{r=2}^\infty \eta_{r}\brk[big]{q_r(u_k)-q_r(u_j)}.
\]
The coefficient $\beta_{r,s}(\lambda)$ is associated to
bilocal deformations $\opbilocal{\opkern{Q}_r}{\opkern{Q}_s}$
composed from two integrable charges $\opkern{Q}_r,\opkern{Q}_s$ \cite{Bargheer:2009xy}.
A new feature as compared to earlier work 
is the coefficient $\eta_r(\lambda)$ 
which originates from bilocal 
deformations $\opbilocal{\opkern{Z}}{\opkern{Q}_r}$ 
involving the spin operator $\opkern{Z}=\half(1-\sigma_z)$.

\paragraph{Charge Eigenvalues.}

It remains to understand the deformations
of the momentum $p(u)$ and the charges $q_r(u)$.
These are induced by the deformation operators $\opboost{Q_r}$ 
together with a suitable rotation of basis to minimize the range. 
The associated parameters are called $\alpha_r(\lambda)$.
Note that the extra boost deformation $\opboost{\opkern{Z}}$
generates the NN twist parameter $\phi(\lambda)$ 
included into the momentum 
(\textit{cf.}\ \eqref{eqn:pandE}).

The functions $p(u)$, $q_r(u)$ are determined by 
differential equations in the parameters $\alpha_r$,
but not in $u$ \cite{Bargheer:2009xy}. 
Therefore they can be written as linear combinations
\begin{align}
p(u) &= 
p^{\text{NN}} (u) +\sum_{r=2}^\infty \gamma_{r}(\lambda) q^{\text{NN}}_r(u),
\nonumber\\
q_r(u) &= q_r^{\text{NN}} (u)+ \sum_{s=r+1}^\infty \gamma_{r,s}(\lambda) q^{\text{NN}}_s(u)
\label{eqn:chargeEV}
\end{align}
of the magnon dispersion relations $p^{\text{NN}}(u)$, $q^{\text{NN}}_r(u)$ 
at leading order (but with the full quantum-deformation parameter $\hbar(\lambda)$).
The dispersion relations for the XXZ spin chain are well-known, 
see  \eqref{eqn:pandE,highercharges}.
The coefficients $\gamma_{r,s}$ are precisely the same as for the XXX system
since the minimization of length turns out to be universal. 
In the XXX system the $\gamma_{r,s}$ can be identified as expansion coefficients 
for the following equations \cite{Bargheer:2009xy}
\begin{align}
\log x(u)
&=
\log u
-
\sum_{s=2}^\infty \gamma_{s}(\lambda) 
\frac{1}{s-1}\,\frac{1}{u^{s-1}}\,,
&
\frac{1}{x(u)^{r-1}}
&=
\sum_{s=r}^\infty \gamma_{r,s}(\lambda) 
\frac{r-1}{s-1}\,\frac{1}{u^{s-1}}\,.
\end{align}
Here the function $x(u)$ is defined as the inverse of 
the function 
\[
u(x):=x+\sum_{n=3}^\infty \frac{\alpha_n}{x^{n-2}}\,.
\]

Note that the above relationships can also be expressed
by means of a generating function of charges and
the residue theorem \cite{Bargheer:2009xy}.


\section{Open Spin Chain}
\label{sec:openchain}

In this section we generalize the above investigations to the open XXZ 
spin chain with diagonal boundary conditions. 
For open integrable spin chains merely the even charges $\opkern{Q}_{2r}$ ($r=1,2,\dots$) 
with range $\range{\opkern{Q}_{2r}}=2r$ furnish conserved quantities 
while the odd charges only commute up to boundary terms. 
Similar long-range models 
with open boundary conditions were studied in  \cite{Beisert:2008cf,Loebbert:2012yd} 
and we will heavily rely on those results. 

\subsection{NN Hamiltonian and Integrability}

Consider the following open spin chain Hamiltonian 
with diagonal boundary conditions 
\begin{align}
\oper{H}_\text{open}&=\opkern{H}_\Left+\oplocal{\opkern{H}_{\text{bulk}}}+\opkern{H}_\Right,
&
\opkern{H}_{\LeftRight} &= h_{1,\LeftRight} + h_{2,\LeftRight} \sigma_z.
\label{eqn:Hopen}
\end{align}
The bulk Hamiltonian $\opkern{H}_{\text{bulk}}$ is the same as in 
\eqref{eqn:HNN} and the boundary Hamiltonians $\opkern{H}_{\LeftRight}$
act on one site at the left or right boundary, respectively.

Also this Hamiltonian is integrable as can be seen using the methods 
introduced by Sklyanin \cite{Sklyanin:1988yz} 
(\textit{cf.}\ also \cite{deVega:1993xi,Mezincescu:1990uf} and \appref{sec:boundscatt}): 
The bulk Hamiltonian follows from the same arguments as above,
thus the coefficients $h_{1,\dots,6}$ are
parametrized as in the periodic case \eqref{eqn:finalcoeff}.
Consider now the diagonal reflection matrices
\begin{align}\label{eqn:Kmatrix}
\opkern{K}_\Left(u)&=\opkern{K}(u, \xi_\Left,\zeta_\Left) =
e^{i\zeta_\Left u}\,\diag\bigbrk{
e^{-i\psi u}\sinh\hbar(\xi_\Left+u),
e^{+i\psi u}\sinh\hbar(\xi_\Left-u)},
\\
\opkern{K}_\Right(u)&=\opkern{K}(-u-i\kappa,-\xi_\Right,0)\opkern{M},
\qquad\qquad\opkern{M}=\diag(e^{\kappa\psi},e^{-\kappa\psi}). \label{eqn:Kmatrixp}
\end{align}
The boundary contributions to the Hamiltonian 
coming from the reflection matrices take the form
\begin{align}
\opkern{H}_\Left
&=-\frac{i}{2}\lreval{\opkern{K}_\Left^{-1}(u)\frac{d\opkern{K}_\Left(u)}{du}}_{u=0}
=-\frac{1}{2}\bigbrk{\psi+i\hbar \coth \hbar \xi_\Left}\sigma_z+\frac{\zeta_\Left}{2},\\
\opkern{H}_\Right
&=\frac{\Tr_0\bigsbrk{\bigbrk{1\otimes \opkern{K}_\Right(0)}\opkern{H}_{L,0}}}{\Tr\,\opkern{K}_\Right(0)}
=+\frac{1}{2}\bigbrk{\psi -i\hbar \coth \hbar \xi_\Right}\sigma_z\nonumber\\
&\qquad\qquad\qquad\qquad\qquad\qquad\quad+\bigbrk{\rho+\ihalf \hbar \coth i\hbar \kappa-\frac{\psi}{2}\coth \hbar \xi_\Right \tanh i\hbar \kappa}.
\end{align}
The complete Hamiltonian \eqref{eqn:Hopen} is thus obtained by identifying
\begin{align}
h_{1,\Left}&=\frac{1}{2}\zeta_{\Left},
&
h_{1,\Right}&=\rho+\ihalf \hbar \coth i\hbar \kappa-\frac{\psi}{2}\coth \hbar \xi_\Right \tanh i\hbar \kappa,\nonumber\\
h_{2,\Left}&=-\frac{1}{2}\bigbrk{\psi+i\hbar \coth\hbar \xi_\Left},
&
h_{2,\Right}&=\frac{1}{2}\bigbrk{\psi -i\hbar \coth \hbar \xi_\Right}. 
\end{align}
Note that the coefficient $\psi$ cancels among the contribution $h_6=-\psi/2$ 
from the R-matrix and the contributions $h_{2,\LeftRight}$ coming from the K-matrices.
As for the closed chain, it thus completely drops out of the 
overall open spin chain Hamiltonian $\oper{H}_\text{open}$
since it parametrizes the similarity transformation \eqref{eqn:PsiSimTrans} 
extended to the K-matrices \cite{Mezincescu:1990uf}
\begin{align}
\opkern{K}_\Left(u) &\to \opkern{G}(u)^{-1} \opkern{K}_\Left(u) \opkern{G}(\bar u),
&
\opkern{K}_\Right(u) &\to \opkern{G}(\bar u)^{-1} \opkern{K}_\Right(u) \opkern{G}(u).
\end{align}

Now we again remove some inessential freedom from the Hamiltonian to keep the analysis 
simple. In particular we eliminate the vacuum energy contribution from the boundary
by adjusting $\zeta_-$ 
such that $h_{1,\Left}+h_{2,\Left}+h_{1,\Right}+h_{2,\Right}=0$. 
As before, we shall also normalize $h_2=\psi=0$ and $\kappa=1$.

The double-row transfer matrix generalizing \eqref{eq:transfer} to open boundaries takes the form
\begin{equation}
\oper{T}_\text{open}(u)=\Tr_a\opkern{K}_{\Right}(u)\opkern{L}_{a,1}(u)\dots\opkern{L}_{a,L}(u)\opkern{K}_{\Left}(u)\opkern{L}_{a,L}^{-1}(\bar u)\dots\opkern{L}_{a,1}^{-1}(\bar u),
\end{equation}
and the Bethe equations for the open NN chain read 
\begin{align}\label{eqn:openBAE}
\left[
\frac{\sinh\hbar(u_k+\ihalf\kappaR) }{\sinh\hbar( u_k-\ihalf\kappaR)}\,
\frac{\sinh\hbar(\bar u_k-\ihalf\kappaR) }{\sinh\hbar(\bar u_k+\ihalf\kappaR)}\right]^L
&=
\frac{\sinh\hbar( \xi_-- u_k-\sfrac{i}{2}\kappaR)}
{\sinh\hbar (\xi_-+ u_k-\sfrac{i}{2}\kappaR)}
\,\frac{\sinh\hbar (\xi_++\bar u_k-\sfrac{i}{2}\kappaR)}
{\sinh\hbar( \xi_+- \bar u_k-\sfrac{i}{2}\kappaR)}
\nonumber\\
&
\qquad\times\mathop{\prod_{j=1}^M}_{j\neq k} 
 \frac{\sinh\hbar(u_k-u_j+i\kappaR) }{\sinh\hbar(u_k-u_j-i\kappaR)}\,
\frac{\sinh\hbar(\bar u_k-u_j-i\kappaR) }{\sinh\hbar(\bar u_k-u_j+i\kappaR)},
\end{align}
where $\bar u=-u$ denotes the reflection map. 
The latter is defined by requiring 
that in the bulk a plane wave with momentum $p$ 
has the same dispersion relation as the reflected plane wave 
with momentum $\bar p$: $\oper{H}\ket{p}=\oper{H}\ket{\bar p}$.  
Note that $e^{i\bar p}= e^{-2 i \phi}e^{-i p}=e^{ip(\bar u)}$.

The open chain Bethe equations are invariant under the reflection map
for each individual Bethe root, $u_j\to \bar u_j=-u_j$.
In that sense, each Bethe root is defined only modulo its sign.

The coefficients $\xi_{\LeftRight}$ parametrize two boundary phases 
which have an independent effect on the spectrum.
The twist parameter $\phi$ does not appear in the Bethe equations
since the associated NN generator $\opboost{\opkern{Z}}$ 
corresponds to a globally defined similarity transformation on the open spin chain.

\subsection{Deformations on Semi-Infinite Chains}

Open long-range spin chains are defined perturbatively 
in analogy to the periodic case. 
On open chains the basis of non-trivial local operators 
is enlarged by the set of \emph{boundary terms} taking the form
\begin{align}\label{eqn:bound}
\opkern{L}_1&=\opkern{L}-1\otimes\opkern{L},
&
\opkern{L}_L&=\opkern{L}-\opkern{L}\otimes 1.
\end{align}
These operators only act on a left or right boundary, respectively, 
and vanish identically on periodic spin chains.
In other words, the equivalence relation \eqref{eq:L1=1L=L} 
is released on the open chain and each representative 
of the periodic equivalence classes becomes a distinct operator.
For instance, this allows to rewrite the contribution 
from the boundary Hamiltonians $\opkern{H}_\LeftRight$ in \eqref{eqn:Hopen} 
as the kernel
\[
\opkern{H}_\text{open} =
(\opkern{H}_\Left-1\otimes\opkern{H}_\Left) 
+\opkern{H}_{\text{bulk}}
+(\opkern{H}_\Right-\opkern{H}_\Right\otimes 1).
\]

Similarly to the wrapping order for periodic chains,
the definition of long-range open chains breaks down at \emph{spanning order},
that is at the perturbative order where the charge 
deformations span the whole spin chain. 
We will implement this by setting the following \emph{spanning terms} 
to zero throughout this chapter:
\begin{equation}
\opkern{L}_{1,L}=1\otimes \opkern{L}\otimes 1-1\otimes \opkern{L}-\opkern{L}\otimes 1+\opkern{L}.
\end{equation}
Spanning terms exclusively act on chains of length $\range{\opkern{L}}$ and vanish on all other chains.

\paragraph{Semi-infinite chains.}

In \cite{Loebbert:2012yd} it was introduced how to obtain long-range deformations 
for finite integrable spin chains with open boundaries from two semi-infinite chains. 
The method is briefly sketched as follows: 
Making use of the deformation equation \eqref{eqn:DefEQN} 
on left- ($\Left$) and right- ($\Right$) open spin chains%
\footnote{A left-open chain is a semi-infinite chain with 
an open boundary on the left hand side and infinite extent on the right hand side. 
A right-open chain is defined analogously.}
\footnote{Note that $\oplocal{\opkern{L}}$ denotes application 
of the kernel $\opkern{L}$ to the whole spin chain 
no matter whether we are dealing with finite or semi-infinite chains.}
\begin{equation}\label{eqn:DefOpen}
\eval[Big]{
\frac{d}{d\lambda}\oplocal{\opkern{Q}_{2r,\LeftRight}(\lambda)}
=i\comm{\oper{X}_{\LeftRight}(\lambda)}{\oplocal{\opkern{Q}_{2r,\LeftRight}(\lambda)}}
}_\LeftRight\,,
\end{equation}
two sets of deformed charges are found. 
Here $|_\LeftRight$ denotes application 
of left- or right-open boundary conditions, respectively. 
We may then define charges $\opkern{Q}_{2r}$ on a finite open chain by requiring that
\begin{align}\label{eqn:combine}
&\eval[big]{\opkern{Q}_{2r}\simeq\opkern{Q}_{2r,\Left}}_\Left,
&
&\eval[big]{\opkern{Q}_{2r}\simeq\opkern{Q}_{2r,\Right}}_\Right.
\end{align}
Note that this implies that the bulk equivalence classes of $\opkern{Q}_{2r,\Left}$ 
and $\opkern{Q}_{2r,\Right}$ have to match each other 
which can be achieved by adaption of the respective deformation coefficients.

In order to obtain local deformations, the generators $\oper{X}_\Left$ and $\oper{X}_\Right$ 
for the left- and right-open case can and have to be chosen differently. 

The above construction allows to use the odd charges of the periodic system 
for building generators of the open long-range model. 
In fact, only generators involving these odd charges 
yield non-trivial long-range deformations. 
The odd charges, however, are only well defined modulo boundary terms, 
\textit{i.e.}\ up to the equivalence relation \eqref{eq:L1=1L=L}. 
Modification of
these boundary terms in a bilocal generator amounts 
to a local deformation that can be compensated by other local generators.

In the regime of validity of open long-range spin chains 
the left and right boundaries decouple.
In what follows we will therefore mostly discuss left-open spin chains explicitly.
This implies that all operators of the form 
$\opkern{L}_L$ in \eqref{eqn:bound} are set to zero.
We may obtain the respective contributions 
for the right boundary using the parity map. 

\paragraph{Brute force deformations.}

In analogy to the periodic case we study the commutator 
of the first two long-range charges \eqref{eqn:pertcommrel} 
at first perturbative order in the expansion parameter $\lambda$
on left-open chains:
\begin{equation}\label{eq:firstorderopen}
\comm[big]{\oplocal{\opkern{Q}_2^{(1)}}}{\oplocal{\opkern{Q}_4^{(0)}}}
+\comm[big]{\oplocal{\opkern{Q}_2^{(0)}}}{\oplocal{\opkern{Q}_4^{(1)}}}=0.
\end{equation}
In total, we find that the solution to this equation has 13 degrees of freedom, 
\textit{i.e.}\ the 10 bulk parameters found above 
plus 3 parameters coming from the boundary.%
\footnote{Note that here and in the following we will neglect trivial 
degrees of freedom where the identity acts on the left ($1_1$) 
or right ($1_L$) boundary, respectively. 
Both are formally encoded by the operator $1-1\otimes 1$.
 The corresponding terms will be set to zero 
to simplify the expressions and the counting of degrees of freedom. 
Reintroducing these terms is required to keep the vacuum eigenvalue 
of the charges trivial.}
 
The homogeneous solutions are given by
\begin{align}
\bar{\opkern{Q}}_2^{(1)}&= a_1 + a_2 \opkern{Z} + a_3 \opkern{Q}_2^{(0)},\\
\bar{\opkern{Q}}_4^{(1)}&= b_1 + b_2 \opkern{Z} + b_3 \opkern{Q}_2^{(0)}+ b_4 \opkern{Q}_4^{(0)},
\end{align}
and correspond to a trivial shift of the spectrum parametrized 
by the free coefficients $a_i$, $b_i$. 
The remaining solutions of \eqref{eq:firstorderopen} yield ten 
independent degrees of freedom $a_{4,\dots,13}$ and the parameters $a_{4,\dots,10}$ 
correspond to the periodic deformations $a_4\opkern{Q}_3^{(0)}$ in \eqref{eqn:trivdeform} 
and  \eqref{eqn:firstorder} in the periodic limit, \textit{i.e.}\ 
when the periodic equivalence relation \eqref{eq:L1=1L=L} is applied. 
The coefficients $a_{11,\dots,13}$ 
represent new degrees of freedom on the semi-infinite chain. 
The explicit operators are lengthy and their form is not very illuminating.
In the following we will express them in terms of deformation generators 
as in the periodic case.%
\footnote{As expected we find $16=10+3+3$ free parameters 
when counting the degrees of freedom on the finite open chain. 
That is to say that the number of pure boundary parameters 
gets doubled on the finite chain.
In addition, there is 1 parameter to adjust the vacuum 
energy due to the presence of boundaries.}

\paragraph{Deformation equation.}

In this paragraph we explicitly discuss the deformations on the left-open chain 
to match the degrees of freedom found in the previous paragraph.
We distinguish the following nontrivial deformation generators:%
\footnote{Note that $\opbilocal{\opkern{\opkern{Q}}_{2r}}{\opkern{\opkern{Q}}_{2s+1}}$ and $\opbilocal{\opkern{Z}}{\opkern{Q}_{2s+1}}$ 
do not contribute at the leading order generated by the brute force method.}

\begin{itemize}
\item odd boosted charges $\opboost{\opkern{Q}_{2r+1}}$,
\item even/odd bilocal charges $\opbilocal{\opkern{Q}_{2r}}{\opkern{Q}_{2s+1}}$ and $\opbilocal{\opkern{Z}}{\opkern{Q}_{2s+1}}$,
\item odd local charges $\opkern{Q}_{2r+1}$.
\end{itemize}
Here the boost operator on the left-open chain 
is defined as $\opboost{\opkern{L}}=\opbilocal{1}{\opkern{L}}$
(on the right-open chain it is $\opbilocal{\opkern{L}}{1}$).  
Furthermore other local operators as well as 
the generators $\opboost{\opkern{Q}_{2r}}$, $\opboost{\opkern{Z}}$ 
and $\opbilocal{\opkern{Q}_{2r,}}{\opkern{Q}_{2s}}$ 
yield similarity transformations of the open spin chain model.

Let us now match the deformations obtained in the previous paragraph 
by considering the deformation equation \eqref{eqn:DefOpen} 
on a left-open spin chain. 
The ten independent degrees of freedom at range three correspond 
to the following deformations mainly induced by \eqref{eqn:DefOpen}:
\begin{align}
&a_4 i\comm[big]{\opboost{\opkern{H}^{(0)}}}{\opkern{H}^{(0)}}
+a_5 \bigbrk{e^{-i\phi}\sigma_+\otimes\sigma_-+e^{i\phi}\sigma_-\otimes\sigma_+}
+a_6i\comm[big]{\opboost{\opkern{Z}}}{\opkern{H}^{(0)}}\nonumber\\
&+a_7(i\comm[big]{\opboost{\opkern{Q}_3^{(0)}}}{\opkern{H}^{(0)}}-\sfrac{3}{2} \opkern{Q}_4^{(0)})
+a_8i\comm[big]{\opbilocal{\opkern{Z}}{\opkern{H}^{(0)}}}{\opkern{H}^{(0)}}
+a_9i\comm[big]{\oplocal{\sigma_+\otimes\sigma_--\sigma_-\otimes\sigma_+}}{\opkern{H}^{(0)}}
\nonumber\\
&+a_{10}i\comm[big]{\oplocal{\sigma_+\otimes\sigma_-+\sigma_-\otimes\sigma_+}}{\opkern{H}^{(0)}}
+a_{11}i\comm[big]{\oplocal{\opkern{Q}_3^{(0)}}}{\opkern{H}^{(0)}}
+a_{12}(\opkern{Z}-1\otimes\opkern{Z})
\nonumber\\
&+a_{13}i\comm[big]{\oplocal{\sigma_z-1\otimes\sigma_z}}{\opkern{H}^{(0)}}.
\end{align}
On the finite open chain, the corresponding deformations have the following physical interpretations: The paramter $a_4$ as well as the twist deformation parameter $a_6$ represent boost similarity transformations. Furthermore $a_8$ denotes a bilocal similarity transformation while $a_9$ and $a_{10}$ correspond to local bulk similarity transformations. The parameter $a_7$ represents a deformation of the rapidity map, \textit{i.e.}\ the map between the rapidity $u$ and the momentum $p$. The coefficients $a_5$ and $a_{12}$ are not captured by the deformation equation and correspond to making the parameters $\hbar$ and $\xi$ functions of the deformation parameter $\lambda$. The parameters $a_{11\dots13}$ multiply pure boundary terms that are discussed in more detail below.
Note that the operator proportional to $a_7$ has range three only 
for a certain choice of boundary terms of $\opkern{Q}_3^{(0)}$.%
\footnote{Alternatively, the terms of range four can be compensated
by a local operator transformation.}
Here we have used
\begin{align}
\opkern{Q}_3^{(0)}  =\mathord{}&\frac{i\hbar^2}{2} \frac{\coth i\hbar \kappaR}{\sinh i\hbar \kappaR}
\Big[\frac{e^{2i\phi}}{\cosh i\hbar \kappaR}\, \sigma_-\otimes\sigma_z\otimes\sigma_+ -
e^{i\phi}\, (\sigma_-\otimes\sigma_+\otimes\sigma_z + \sigma_z\otimes\sigma_-\otimes\sigma_+)
\nonumber\\
&+e^{-i\phi}\, (\sigma_+\otimes\sigma_-\otimes\sigma_z +  \sigma_z\otimes\sigma_+\otimes\sigma_-)
- \frac{e^{-2i\phi}}{\cosh i\hbar \kappaR}\, \sigma_+\otimes\sigma_z\otimes\sigma_-\Big]
\nonumber\\
&+\frac{i\hbar^2}{2}\Big[\coth \hbar \xi_-\bigbrk{e^{i\phi}(1\otimes\sigma_-\otimes\sigma_+-\sigma_-\otimes\sigma_+)
-e^{-i\phi}(1\otimes\sigma_+\otimes\sigma_--\sigma_+\otimes\sigma_-)}
\nonumber\\
&+\coth \hbar \xi_+\bigbrk{e^{i\phi}(\sigma_-\otimes\sigma_+\otimes 1-\sigma_-\otimes\sigma_+)
-e^{-i\phi}(\sigma_+\otimes\sigma_-\otimes1-\sigma_+\otimes\sigma_-)}\Big].
\end{align}

\paragraph{Boundary terms on finite chains.}

In order to promote the above deformations to the finite open chain one may 
follow the procedure outlined in \cite{Loebbert:2012yd} 
and combine the deformations 
from the left-open chain with corresponding deformations on the 
right-open chain according to \eqref{eqn:combine}. 
The combined deformations are constructed to have matching bulk terms,
\textit{e.g.}\ $a_{k}^{\Left}=a_{k}^{\Right}$ for $k=1,\ldots,10$,
which guarantees integrability of the resulting charges.
Conversely, pure boundary contributions can be different for the two semi-infinite chains,
\textit{e.g.}\ the left-open parameters
$a_{k}^{\Left}$ for $k=11,\ldots,13$
are independent of the corresponding right-open parameters $a_{k}^{\Right}$.

Note the special role of the odd charges $\opkern{Q}_{2r+1}$ 
as a deformation operator in this context:
Since the latter generate pure boundary terms 
(\textit{i.e.}\ the bulk terms trivially match), 
the deformations on the left- and right-open chains 
yield admissible independent long-range deformations.
We thus obtain two degrees of freedom $a_{11}^{\Left}, a_{11}^{\Right}$ on the finite chain. 
The sum of these degrees of freedom corresponds to a similarity transformation 
while the difference generates a boundary phase. 

Also the parameters $a_{12}$ and $a_{13}$
each split into a left and right boundary degree of freedom on the finite chain. 
The parameters $a_{13}^{\Left}$ and $a_{13}^{\Right}$ 
both represent similarity transformations 
induced by boundary operators
while the coefficients $a_{12}^{\Left}$ and $a_{12}^{\Right}$ 
correspond to making the NN boundary parameters $\xi_\LeftRight$ dependent on $\lambda$.

\subsection{Asymptotic Spectrum}

In \cite{Loebbert:2012yd} it was shown that the long-range Bethe equations for $M$
magnons on an open spin chain of length $L$  take the form
(where $\bar u=-u$ denotes the rapidity of reflected magnons)
\begin{equation}\label{eq:openBethe}
\exp\bigbrk{ip(u_k)L-i\bar p(u_k)L}=
S_{\Left}(u_k,\bar u_k)S_{\Right}(\bar u_k,u_k)\mathop{\prod_{j=1}^M}_{j\neq k} S(u_k,u_j)S^{-1}(\bar u_k,u_j).
\end{equation}
The dispersion relations $p(u)$, $q_{2r}(u)$ take the form \eqref{eqn:chargeEV} 
with the inessential parameters $\phi=\alpha_{2r}=\gamma_{2r,2s+1}=0$.%
\footnote{As discussed above, all of these parameters correspond to 
globally-defined similarity transformations which have no impact on the spectrum.
One might leave these parameters in the Bethe equations and make sure that they 
have no effect under the transformation $u\to\bar u$. 
However, such a cancellation appears to require substantially more complicated expressions.
To avoid such (unphysical) complications, 
we shall simply set these unessential parameters to zero.
Consequently, all charges are parity-even: 
$\bar u=-u$, $\bar p(u)=-p(u)$, $\bar q_{r}=(-1)^r q_r$.}
The scattering factors are given by the bulk expression
\begin{equation}
S(u,u')=\frac{\sinh\hbar(u-u'+i\kappaR)}
{\sinh\hbar(u-u'-i\kappaR)}\,e^{2i\theta(u,u')}
\end{equation}
and the boundary factor
\begin{equation}
S_{\Left}(u,\bar u)S_\Right(\bar u,u)
=\frac{\sinh\hbar( \xi_{\Left}- u-\sfrac{i}{2}\kappaR)\sinh\hbar( \xi_++\bar u-\sfrac{i}{2}\kappaR)}
{\sinh\hbar( \xi_{\Left}+ u-\sfrac{i}{2}\kappaR)\sinh\hbar( \xi_{\Right}-\bar u-\sfrac{i}{2}\kappaR)}\,
e^{2i \omega_{\Left}(u,\bar u)+2i\omega_{\Right}(\bar u,u)}.
\end{equation}
The dressing phase $\theta$ takes the form \eqref{eqn:dressphase}
with $\beta_{2r+1,2s+1}=\beta_{2r,2s}=\eta_{2r}=0$ 
and the reflection phase $\omega(u,\bar u)$ is given by%
\footnote{Note that the dressing phase term $\theta(\bar u,u)$ in the boundary phase can be understood as a contribution from the product on the right hand side of \eqref{eq:openBethe} for $j=k$.}
\begin{equation}\label{eqn:boundphase}
\omega_{\Left}(u,\bar u)+\omega_{\Right}(\bar u, u)=\sum_{r=1}^\infty \delta_{2r+1} q_{2r+1}(u)-\theta(\bar u, u).
\end{equation}
Here the parameters $\delta_{2r+1}$ are induced by deformations with the odd charges. 
For details on the boundary phase induced by bilocal charges see \appref{sec:overlap}.
The individual left and right boundary reflection factors $S_\Left$ and $S_\Right$
receive additional deformations from $\opboost{\opkern{Q}_{2r+1}}$, $\opbilocal{\opkern{Z}}{\opkern{Q}_{2r+1}}$ 
and $\opbilocal{\opkern{Q}_{2r}}{\opkern{Q}_{2s+1}}$. 
The physical quantity appearing in the Bethe equations is the product $S_\Left(u) S_\Right(\bar u)$ 
which does not contain these deformations. 


\section{Conclusions}
\label{sec:conclusions}

\begin{table}\centering
\begin{tabular}{|l|l|l|}
\hline
Operator&Deformation&Parameter\\\hline
$\oplocal{\opkern{Q}_{r}}$&trivial&---\\
$\oplocal{\opkern{L}}$&similarity&$[\epsilon]$\\
$\opboost{\opkern{Q}_{r}}$&rapidity&$\alpha_{r}$\\
$\opbilocal{\opkern{Q}_{r}}{\opkern{Q}_{s}}$&dressing phase &$\beta_{r,s}$\\
$\opboost{\opkern{Z}}$&twist&$\phi$\\
$\opbilocal{\opkern{Z}}{\opkern{Q}_{r}}$&momentum twist&$\eta_{r}$\\
---&q-deformation&$\hbar$\\
basis change&eigenvalues&$\gamma_{r,s}$, $\gamma_{r,1}$, $\gamma_{r,\oper{Z}}$ \\
\hline
\end{tabular}
\caption{Overview of different integrable deformations for the XXZ spin 
chain with periodic boundary conditions. The parameters in square brackets have no impact on the spectrum.}
\label{tab:overviewclosed}
\end{table}

In this paper we considered the most general nearest-neighbor spin 
chain Hamiltonian which preserves the number of up and down spins. 
This underlying nearest neighbor spin chain is equivalent 
to the XXZ spin chain with some additional twist parameters. 
We determined its spectrum by means of the algebraic Bethe ansatz. 

We then classified all long-range deformations of this Hamiltonian 
that preserve integrability. 
This was first done by explicitly computing all such deformations 
of the Hamiltonian of range three and four or less. 
We found that the Hamiltonian admits ten possible deformations 
of at most range three \eqref{eqn:trivdeform,eqn:firstorder} 
and twenty-three at range four. They can be split into three parts; 
trivial, NN and long-range. The trivial deformations \eqref{eqn:trivdeform} 
simply correspond to adding operators that already 
commute with the conserved charges. 
The NN deformations are those that arise from a change in the parameters 
of the Hamiltonian. 
This then leaves four non-trivial long-range deformations described 
by the terms $a_{7,8,9,10}$ from \eqref{eqn:firstorder}.

The found deformations were subsequently understood in the general framework 
put forward in \cite{Bargheer:2009xy,Bargheer:2008jt}. 
This formalism uses the first order equation \eqref{eqn:DefEQN} 
to generate deformations. We showed that all long-range deformations 
are expressible by suitable operators in the deformation equation. 
We also found that one of the NN deformations, the twist, 
is generated by the boost of the spin operator 
in z-direction $\opboost{\opkern{Z}}$.
Unfortunately, we could not fit the q-deformation parameter $\hbar$
into the general deformation framework. 
We did not find a deformation operator that would
generate the quantum deformation consistently for all generators. 
This result is in fact to be expected, 
since $\hbar$ parametrizes a family of
inequivalent quantum groups whereas the deformation framework
does not deform the underlying quantum group.
It would be interesting to find out 
if and how the framework could be generalized 
to accommodate this deformation parameter.
It would, in other words, allow to deform the XXX chain into 
the XXZ chain.

The effect of the long-range deformations on the Bethe equations 
and corresponding spectrum was also derived. 
It affects the spectrum by introducing 
a so-called dressing phase \eqref{eqn:dressphase} 
and by modifying the dispersion relation \eqref{eqn:chargeEV}. 

Finally we studied the open XXZ spin chain. 
We considered the case of diagonal boundary conditions. 
We again determined the possible deformations 
by a brute force calculation and matched those to solutions 
of the deformation equation. 
Here a distinction has to be made between even and odd charges.
All possible deformations and their effect on the spectrum 
for both open and closed spin chains are summarized in \tabref{tab:overviewclosed}
and \tabref{tab:overviewopen}.

\begin{table}\centering
\begin{tabular}{|l|l|l|l|}
\hline
Left-open& Right-open&Deformation&Parameter\\\hline
$\oplocal{\opkern{Q}_{2r+1}}$&$+\oplocal{\opkern{Q}_{2r+1}}$&similarity&$[\epsilon]$\\
$\oplocal{\opkern{Q}_{2r+1}}$&$-\oplocal{\opkern{Q}_{2r+1}}$&boundary phase&$\delta_{2r+1}$\\
$\oplocal{\opkern{Q}_{2r}}$&$\pm\oplocal{\opkern{Q}_{2r}}$&trivial&---\\
$\oplocal{\opkern{L}}$&$+\oplocal{\opkern{L}}$&similarity&$[\epsilon]$\\
$\opbilocal{1}{\opkern{Q}_{2r+1}}$&$-\opbilocal{\opkern{Q}_{2r+1}}{1}$&rapidity&$\alpha_{2r+1}$\\
$\opbilocal{1}{\opkern{Q}_{2r}}$&$+\opbilocal{\opkern{Q}_{2r}}{1}$&similarity&[$\alpha_{2r}$]\\
$\opbilocal{\opkern{Q}_{2r}}{\opkern{Q}_{2s+1}}$&$-\opbilocal{\opkern{Q}_{2s+1}}{\opkern{Q}_{2r}}$&dressing phase &$\beta_{2r,2s+1}$\\
$\opbilocal{\opkern{Q}_{2r}}{\opkern{Q}_{2s}}$&$+\opbilocal{\opkern{Q}_{2s}}{\opkern{Q}_{2r}}$&similarity&[$\beta_{2r,2s}$]\\
$\opbilocal{\opkern{Q}_{2r+1}}{\opkern{Q}_{2s+1}}$&$\pm\opbilocal{\opkern{Q}_{2s+1}}{\opkern{Q}_{2r+1}}$&nonlocal&---\\
$\opbilocal{1}{\opkern{Z}}$&$+\opbilocal{\opkern{Z}}{1}$&similarity&[$\phi$]\\
$\opbilocal{\opkern{Z}}{\opkern{Q}_{2r+1}}$&$-\opbilocal{\opkern{Q}_{2r+1}}{\opkern{Z}}$&momentum twist&$\eta_{2r+1}$\\
$\opbilocal{\opkern{Z}}{\opkern{Q}_{2r}}$&$+\opbilocal{\opkern{Q}_{2r}}{\opkern{Z}}$&similarity&[$\eta_{2r}$]\\
---&---&q-deformation&$\hbar$\\
---&---&NN boundary phases&$\xi_\LeftRight$\\
basis change&basis change&eigenvalues&$\gamma_{2r,2s}$, $\gamma_{2r,1}$, $\gamma_{2r,\oper{Z}}$ \\
\hline
\end{tabular}

\caption{Overview of different integrable deformations for the XXZ spin 
chain with open boundary conditions.  The parameters in square brackets have no impact on the spectrum.}
\label{tab:overviewopen}
\end{table}

In this and the preceding studies 
of the considered deformation method, 
several combinations of bilocal charges were identified 
that lead to integrable deformations. 
Recently a dressing phase contribution including 
the momentum was analyzed in the context 
of the $\text{AdS}_3/\text{CFT}_2$ correspondence \cite{Beccaria:2012kb,Borsato:2013hoa}. 
It would be interesting to investigate a bilocal generator $\opbilocal{\opkern{P}}{\opkern{Q}_r}$ 
using this method of deformation. The momentum operator $\opkern{P}$ might also be a hint on the origin of the 
NN boundary phase whose leading contribution is linear in the momentum as opposed to the long-range boundary phases.

\paragraph{Acknowledgements.}
The work of NB and MdL is partially supported by grant no.\ 200021-137616 from the Swiss
National Science Foundation.


\appendix
\addcontentsline{toc}{section}{Appendix} 
\addtocontents{toc}{\protect\setcounter{tocdepth}{-1}}

\section{Solutions of the Yang--Baxter Equation}

The R-matrix used to describe the integrable structure of our Hamiltonian
was found  by explicitly solving the Yang--Baxter equation.
In fact, assuming the R-matrix to be of difference form allows we are able
to find all solutions of the Yang--Baxter equation that are $\alg{gl}(1)$ invariant.

Invariance under $\alg{gl}(1)$ and assuming it to be of difference form
simply implies that the R-matrix can be written as
\begin{align}
R(u) = \begin{pmatrix}
1 & 0 & 0 & 0\\
0 & a(u) & b(u) & 0\\
0 & c(u) & d(u) & 0\\
0 & 0 & 0 & e(u) \\
\end{pmatrix},
\end{align}
where $a,b,c,d,e$ are smooth functions.
As per usual we insist that at $u=0$ the R-matrix reduces to the permutation operator.
This fixes
\begin{align}
&a(0)=d(0)=0, && b(0)=c(0)=e(0)=1.
\end{align}
Suppose that $R$ satisfies the Yang--Baxter equation
\begin{align}
R_{12}(u_1-u_2)R_{13}(u_1-u_3)R_{23}(u_2-u_3) =
R_{23}(u_2-u_3)R_{13}(u_1-u_3)R_{12}(u_1-u_2),
\end{align}
then we find functional relations between the functions $a,b,c,d,e$.
For instance we have
\begin{align}
b(u_1-u_2)b(u_2-u_3)c(u_1-u_3) = b(u_1-u_3)c(u_1-u_2)c(u_2-u_3).
\end{align}
Expanding this around $u_2 = u + u_3$ implies the following differential equation
\begin{align}
(b^\prime(0)-c^\prime(0)) b(u)c(u) + b(u)c^\prime(u) = c(u)b^\prime(u).
\end{align}
Hence, using that $b(0)=c(0)$ this is solved to give
\begin{align}
c(u) = e^{(c^\prime(0)-b^\prime(0))}b(u).
\end{align}
Using the same expansion on all the coefficients of the Yang--Baxter equation we furthermore derive
\begin{align}
&a(u) = \frac{b^\prime(u)}{d^\prime(0) b(u)} -\frac{b^\prime(0)}{d^\prime(0)},
&& b(u)(a^\prime(0) + c^\prime(0)a(u)-a^\prime(u)) + a(u)b^\prime(u)=0,
\end{align}
which are then readily combined to yield a second order differential equation for $a$ only.
In this way we can solve for all the functions in the R-matrix.
We find that a solution only exists if $e^\prime(0) = 0$ or $e^\prime(0) = b^\prime(0)+c^\prime(0)$.
Finally imposing that the R-matrix is unitary,
\textit{i.e.}\ $R_{12}(u)R_{21}(u) = 1$ we find two different solutions%
\footnote{The exact relation between the parameters $\phi,\psi,\hbar,\kappaR$
and the constants from the differential equation is rather involved but can be worked out explicitly},
namely \eqref{eqn:Rmatrix} and
\begin{align}
\tilde{R}(u) =
\begin{pmatrix}
1 & 0 & 0 & 0 \\
0 & e^{\phi}\frac{\sinh\hbar u  }{\sinh\hbar(u + i\kappaR)} &
e^{i\psi u}\frac{\sin\hbar\kappaR}{\sinh\hbar(u + i\kappaR)}  & 0 \\
0 & e^{-i\psi u}\frac{\sin\hbar\kappaR}{\sinh\hbar(u + i\kappaR)} &
e^{-\phi}\frac{\sinh\hbar u}{\sinh\hbar(u + i\kappaR)} & 0 \\
0 & 0 & 0 & -\frac{\sinh\hbar (u-i\kappaR)}{\sinh\hbar(u + i\kappaR)}.
\end{pmatrix}.
\end{align}
These two solutions were expected since they correspond to twisted,
quantum deformed $\alg{sl}(2)$ and $\alg{sl}(1|1)$ respectively.
Nevertheless, we find that these are all solutions of this type.
The R-matrix $\tilde{R}$ generates a Hamiltonian which does not
contain the term $\sigma_z\otimes\sigma_z$ and hence is not suited
to describe the integrable structure of the Hamiltonian \eqref{eqn:HNN}.

\section{Boundary Scattering}
\label{sec:boundscatt}
Let us briefly specify some of the properties of the R- and K-matrices in the context of open spin chain integrability.
The shift $i\kappa$ and the matrix $\opkern{M}$ in \eqref{eqn:Kmatrixp}
are distinguished by the following \emph{crossing unitarity} 
relation satisfied by the R-matrix \eqref{eqn:Rmatrix}:
\begin{equation}
\opkern{R}_{12}^{t_1}(u,\phi)(\opkern{M}\otimes 1)\opkern{R}_{12}^{t_2}(-u-2i\kappa,-\phi)
(\opkern{M}^{-1}\otimes 1)=\zeta(u+i\kappa).
\end{equation}
Here 
\begin{equation}
\zeta(u)=e^{2\kappa\rho}\brk[big]{\cosh^2{i\hbar\kappa}+\coth^2{\hbar u}\sinh^2{i\hbar \kappa}}
\end{equation}
and $t_k$ denotes transposition in the space $k$. 
The R- and K-matrices satisfy the following boundary Yang--Baxter equations:%
\footnote{Note that $\opkern{R}^{-1}(u,\phi)=\opkern{R}^{t_1t_2}(-u,-\phi)$.}
\begin{align}
&\opkern{R}(u-v)\bigbrk{\opkern{K}_\Left(u)\otimes 1}\opkern{R}^{-1}(-u-v)\bigbrk{1\otimes \opkern{K}_\Left(v)}\nonumber\\
&=\bigbrk{1\otimes \opkern{K}_\Left(v)}\opkern{R}(u+v)\bigbrk{\opkern{K}_\Left(u)\otimes 1}\opkern{R}^{-1}(-u+v),\\
&\opkern{R}(-u+v)\bigbrk{\opkern{K}_\Right^{t_1}(u)\opkern{M}^{-1}\otimes 1}\opkern{R}^{-1}(u+v+2i\kappa)
  \bigbrk{\opkern{M}\otimes \opkern{K}_\Right^{t_2}(v)}\nonumber\\
&=\bigbrk{\opkern{M}\otimes \opkern{K}_\Right^{t_2}(v)}\opkern{R}(-u-v-2i\kappa)\bigbrk{\opkern{M}^{-1}\opkern{K}_\Right^{t_1}(u)\otimes 1}\opkern{R}^{-1}(u-v).
\end{align}


\section{Boundary Phase from Bilocal Charges}
\label{sec:overlap}

Consider two bilocal charges $\opbilocal{\opkern{Q}_{2r,\Left}}{\opkern{Q}_{2s+1,\Left}}$ and $\opbilocal{\opkern{Q}_{2s+1,\Right}}{\opkern{Q}_{2r,\Right}}$ representing generators on the left- and right-open spin chain, respectively. Here we assume that the local charges $\opkern{Q}_{2r,\LeftRight}$ and $\opkern{Q}_{2s+1,\LeftRight}$ have vanishing vacuum eigenvalues and that the bilocal charges have the following one-magnon dispersion relation in the bulk of the chain:
\begin{align}
\opbilocal{\opkern{Q}_{2r,\Left}}{\opkern{Q}_{2s+1,\Left}}\ket{u}&=\phi_1(u)\ket{u},
&
\opbilocal{\opkern{Q}_{2s+1,\Right}}{\opkern{Q}_{2r,\Right}}\ket{u}&=\phi_2(u)\ket{u}.
\end{align}
Note that the odd charges $\opkern{Q}_{2s+1,\Left}$ and $\opkern{Q}_{2s+1,\Right}$ may differ by boundary terms but coincide in the bulk. The even charges $\opkern{Q}_{2r,\Left}$ and $\opkern{Q}_{2r,\Right}$ might also differ  by a boundary similarity transformation. Up to a local regularization $\opkern{Y}$, the sum of bilocal charges equals the ordinary product of the local charges in the bulk:
\begin{equation}\label{eq:uptolocal}
\opbilocal{\opkern{Q}_{2r,\Left}}{\opkern{Q}_{2s+1,\Left}}+\opbilocal{\opkern{Q}_{2s+1,\Right}}{\opkern{Q}_{2r,\Right}}\simeq\oplocal{\opkern{Q}_{2r}}\oplocal{\opkern{Q}_{2s+1}}+\oplocal{\opkern{Y}}.
\end{equation}
This implies for the dispersion relations
\begin{equation}
\phi_1(u)+\phi_2(u)=q_{2r}(u)q_{2s+1}(u)+y(u),
\end{equation}
where $y(u)$ denotes the one-magnon dispersion relation of $\oplocal{\opkern{Y}}$. 

Now we may deform with the operators $2i\opbilocal{\opkern{Q}_{2r,\Left}}{\opkern{Q}_{2s+1,\Left}}$ and $-2i\opbilocal{\opkern{Q}_{2s+1,\Right}}{\opkern{Q}_{2r,\Right}}$ which induces boundary scattering phases at the left or right boundary, respectively. These take the form
\begin{align}
\omega_\Left(u,\bar u)&=2i\bigbrk{\phi_1(\bar u)-\phi_1(u)},
&
\omega_\Right(\bar u, u)&=2i\bigbrk{\phi_2(\bar u)-\phi_2(u)}.
\end{align}
The resulting scattering factor in the Bethe equations of the finite open spin chain is then given by
\begin{equation}
\omega_\mathrm{bound}(u,\bar u)=\omega_\Left(u,\bar u)+\omega_\Right(\bar u,u)=-4iq_{2r}(u)q_{2s+1}(u)+y(\bar u)-y(u),
\end{equation}
where we have used $q_{2s+1}(-u)=-q_{2s+1}(u)$. In consequence we find a dressing phase contribution to the boundary phase as well as the parity odd function $y(\bar u)-y(u)$. The latter originates from local operators and can thus be expanded in terms of (finitely many) odd dispersion functions $q_{2n+1}(u)$; hence we may absorb it into the parameters $\delta_{2r+1}$ in \eqref{eqn:boundphase}.  

The appearance of the odd local charges can also be seen from \eqref{eq:uptolocal} by taking into account \eqref{eqn:combine}: For $\opbilocal{\opkern{Q}_{2r,\Left}}{\opkern{Q}_{2s+1,\Left}}$ and $-\opbilocal{\opkern{Q}_{2s+1,\Right}}{\opkern{Q}_{2r,\Right}}$ to induce the same bulk structure, the operator $\opkern{Y}$ is constrained to be a linear combination of $1$, $\opkern{Z}$ and the local charges. Here $1$, $\opkern{Z}$ and the even local charges induce trivial deformations; consistently their eigenvalues drop out of the combination $y(\bar u)-y(u)$.

Note that the above arguments do not allow to compute $\omega_\Left$ and $\omega_\Right$ individually. Correspondingly, when these quantities 
are computed from a single boundary scattering problem in practice, one obtains terms which are hard to interpret. Gladly only their sum appears in the Bethe equations and represents a physical quantity relevant for the spectrum.

\begin{bibtex}{\jobname}
@article{Beisert:2010jr,
      author         = "Beisert, Niklas and others",
      title          = "{Review of AdS/CFT Integrability: An Overview}",
      journal        = "Lett.Math.Phys.",
      volume         = "99",
      pages          = "3-32",
      doi            = "10.1007/s11005-011-0529-2",
      year           = "2012",
      eprint         = "1012.3982",
      archivePrefix  = "arXiv",
      primaryClass   = "hep-th",
      reportNumber   = "AEI-2010-175, CERN-PH-TH-2010-306, HU-EP-10-87,
                        HU-MATH-2010-22, KCL-MTH-10-10, UMTG-270, UUITP-41-10",
      SLACcitation   = "
}

@Article{Haldane:1988gg,
     author    = "Haldane, F. D. M.",
     title     = "Exact Jastrow-Gutzwiller resonating valence bond ground
                  state of the spin $1/2$ antiferromagnetic Heisenberg chain
                  with $1/r^2$ exchange",
     journal   = "Phys. Rev. Lett.",
     volume    = "60",
     year      = "1988",
     pages     = "635",
     doi       = "10.1103/PhysRevLett.60.635",
     SLACcitation  = "
}

@Article{Inozemtsev:1989yq,
     author    = "Inozemtsev, V. I.",
     title     = "On the connection between the one-dimensional s = $1/2$
                  Heisenberg chain and Haldane Shastry model",
     journal   = "J. Stat. Phys.",
     volume    = "59",
     year      = "1990",
     pages     = "1143",
     doi = "10.1007/BF01334745",
}

@Article{Inozemtsev:2002vb,
     author    = "Inozemtsev, V. I.",
     title     = "Integrable Heisenberg-van Vleck chains with variable range
                  exchange",
     journal   = "Phys. Part. Nucl.",
     volume    = "34",
     year      = "2003",
     pages     = "166-193",
     eprint    = "hep-th/0201001",
     SLACcitation  = "
}

@article{Beisert:2008cf,
      author         = "Beisert, N. and Loebbert, F.",
      title          = "{Open Perturbatively Long-Range Integrable gl(N) Spin
                        Chains}",
      journal        = "Adv.Sci.Lett.",
      volume         = "2",
      pages          = "261-269",
      doi            = "10.1166/asl.2009.1034",
      year           = "2009",
      eprint         = "0805.3260",
      archivePrefix  = "arXiv",
      primaryClass   = "hep-th",
      reportNumber   = "AEI-2008-032",
      SLACcitation   = "
}

@article{Beisert:2005wv,
      author         = "Beisert, N. and Klose, T.",
      title          = "{Long-range gl(n) integrable spin chains and plane-wave
                        matrix theory}",
      journal        = "J.Stat.Mech.",
      volume         = "0607",
      pages          = "P07006",
      doi            = "10.1088/1742-5468/2006/07/P07006",
      year           = "2006",
      eprint         = "hep-th/0510124",
      archivePrefix  = "arXiv",
      primaryClass   = "hep-th",
      reportNumber   = "AEI-2005-157, PUTP-2178, UUTTP-19-05",
      SLACcitation   = "
}

@Article{SriramShastry:1988gh,
     author    = "Shastry, B. S.",
     title     = "Exact solution of an S = $1/2$ Heisenberg antiferromagnetic
                  chain with long ranged interactions",
     journal   = "Phys. Rev. Lett.",
     volume    = "60",
     year      = "1988",
     pages     = "639",
     doi       = "10.1103/PhysRevLett.60.639",
     SLACcitation  = "
}

@article{Loebbert:2012yd,
      author         = "Loebbert, Florian",
      title          = "{Recursion Relations for Long-Range Integrable Spin
                        Chains with Open Boundary Conditions}",
      journal        = "Phys.Rev.",
      volume         = "D85",
      pages          = "086008",
      doi            = "10.1103/PhysRevD.85.086008",
      year           = "2012",
      eprint         = "1201.0888",
      archivePrefix  = "arXiv",
      primaryClass   = "hep-th",
      reportNumber   = "LPT-ENS-12-01, AEI-2012-000",
      SLACcitation   = "
}

@article{Bargheer:2008jt,
      author         = "Bargheer, Till and Beisert, Niklas and Loebbert, Florian",
      title          = "{Boosting Nearest-Neighbour to Long-Range Integrable Spin
                        Chains}",
      journal        = "J.Stat.Mech.",
      volume         = "0811",
      pages          = "L11001",
      doi            = "10.1088/1742-5468/2008/11/L11001",
      year           = "2008",
      eprint         = "0807.5081",
      archivePrefix  = "arXiv",
      primaryClass   = "hep-th",
      reportNumber   = "AEI-2008-052",
      SLACcitation   = "
}

@article{Bargheer:2009xy,
      author         = "Bargheer, Till and Beisert, Niklas and Loebbert, Florian",
      title          = "{Long-Range Deformations for Integrable Spin Chains}",
      journal        = "J.Phys.",
      volume         = "A42",
      pages          = "285205",
      doi            = "10.1088/1751-8113/42/28/285205",
      year           = "2009",
      eprint         = "0902.0956",
      archivePrefix  = "arXiv",
      primaryClass   = "hep-th",
      reportNumber   = "AEI-2009-009",
      SLACcitation   = "
}

@article{Sklyanin:1988yz,
      author         = "Sklyanin, E.K.",
      title          = "{Boundary Conditions for Integrable Quantum Systems}",
      journal        = "J.Phys.",
      volume         = "A21",
      pages          = "2375",
      doi            = "10.1088/0305-4470/21/10/015",
      year           = "1988",
      reportNumber   = "E-11-86",
      SLACcitation   = "
}

@article{deVega:1993xi,
      author         = "de Vega, H.J. and Gonz\'alez-Ruiz, A.",
      title          = "{Boundary K matrices for the XYZ, XXZ and XXX spin
                        chains}",
      journal        = "J.Phys.",
      volume         = "A27",
      pages          = "6129-6138",
      doi            ="10.1088/0305-4470/27/18/021",
      year           = "1994",
      eprint         = "hep-th/9306089",
      archivePrefix  = "arXiv",
      primaryClass   = "hep-th",
      reportNumber   = "PAR-LPTHE-93-29, LPTHE-PAR-93-29",
      SLACcitation   = "
}

@article{Mezincescu:1990uf,
      author         = "Mezincescu, Luca and Nepomechie, Rafael I.",
      title          = "{Integrable open spin chains with nonsymmetric R
                        matrices}",
      journal        = "J.Phys.",
      volume         = "A24",
      pages          = "L17-L24",
     doi            ="10.1088/0305-4470/24/1/005",
      year           = "1991",
      reportNumber   = "UMTG-157",
      SLACcitation   = "
}

@Article{Tetelman:1981xx,
     author    = "M.G. Tetelman",
     title     = "{Lorentz group for two-dimensional integrable lattice systems}",
     journal   = "Sov. Phys. JETP.",
     volume    = "55",
     year      = "1982",
     pages     = "306",
     SLACcitation  = "
}
@article{ReshTwist,
year={1990},
issn={0377-9017},
journal={Lett.Math.Phys.},
volume={20},
number={4},
doi={10.1007/BF00626530},
title={Multiparameter quantum groups and twisted quasitriangular Hopf algebras},
publisher={Kluwer Academic Publishers},
keywords={16A24},
author={Reshetikhin, N.},
pages={331-335},
language={English}
}

@article{Faddeev:1996iy,
      author         = "Faddeev, L.D.",
      title          = "{How algebraic Bethe ansatz works for integrable model}",
      pages          = "pp. 149-219",
      year           = "1996",
      eprint         = "hep-th/9605187",
      archivePrefix  = "arXiv",
      primaryClass   = "hep-th",
      SLACcitation   = "
}

@article{Beccaria:2012kb,
      author         = "Beccaria, M. and Levkovich-Maslyuk, F. and Macorini, G.
                        and Tseytlin, A.A.",
      title          = "{Quantum corrections to spinning superstrings in $\text{AdS}_3 \times
                        \text{S}^3 \times \text{M}^4$: determining the dressing phase}",
      journal        = "JHEP",
      volume         = "1304",
      pages          = "006",
      doi            = "10.1007/JHEP04(2013)006",
      year           = "2013",
      eprint         = "1211.6090",
      archivePrefix  = "arXiv",
      primaryClass   = "hep-th",
      reportNumber   = "IMPERIAL-TP-AAT-2012-07, ITEP-TH-56-12",
      SLACcitation   = "
}

@article{Borsato:2013hoa,
      author         = "Borsato, Riccardo and Sax, Olof Ohlsson and Sfondrini,
                        Alessandro and Stefa\'nski,~jr., Bogdan and Torrielli,
                        Alessandro",
      title          = "{Dressing phases of $\text{AdS}_3/\text{CFT}_2$}",
      journal        = "Phys.Rev.",
      volume         = "D88",
      pages          = "066004",
      doi            = "10.1103/PhysRevD.88.066004",
      year           = "2013",
      eprint         = "1306.2512",
      archivePrefix  = "arXiv",
      primaryClass   = "hep-th",
      reportNumber   = "DMUS-MP-13-14\\-ITP-UU-13-14\\-SPIN-13-10-DMUS-MP-13-14,
                        --ITP-UU-13-14, SPIN-13-10, DMUS-MP-13-14, ITP-UU-13-14",
      SLACcitation   = "
}

\end{bibtex}

\bibliographystyle{nb}
\bibliography{\jobname}

\end{document}